\def\ninabla{\hat{n}_1\cdot\nabla}
\def\njnabla{\hat{n}_2\cdot\nabla}
\def\nknabla{\hat{n}_3\cdot\nabla}
\def\niig{\hat{n}_1\cdot\hat\nabla}
\def\njjg{\hat{n}_2\cdot\hat\nabla}
\def\nkkg{\hat{n}_3\cdot\hat\nabla}
\def\Pr{\mbox{\rm Pr}}

\documentstyle[epsf]{elsart}

\newcommand{\be}{\begin{equation}}
\newcommand{\ee}{\end{equation}}
\newcommand{\bea}{\begin{eqnarray}}
\newcommand{\eea}{\end{eqnarray}}

\input{epsf}

\begin{document}

\begin{frontmatter}



\title{Mean flow in hexagonal convection: stability and nonlinear dynamics}


\author{Yuan-nan Young\thanksref{Co}}
\author{\and Hermann Riecke}

\address{
Department of Engineering Sciences and Applied Mathematics, 
Northwestern University,2145 Sheridan Rd, Evanston, IL, 60208, USA}

\thanks[Co]{Corresponding author. 
Tel.:(847) 467-3345, Fax:(847) 491-2178.
email: young@statler.esam.nwu.edu}

\begin{abstract}

Weakly nonlinear hexagon convection patterns coupled to mean flow are
investigated  within the framework of coupled Ginzburg-Landau equations.
The equations are in particular relevant for non-Boussinesq
Rayleigh-B\'enard  convection at low Prandtl numbers. The mean flow is
found to (1) affect only one of the two long-wave phase modes of the
hexagons and (2) suppress the mixing between the two phase modes. 
As a consequence, for small Prandtl numbers the transverse and
the longitudinal phase instability occur in sufficiently distinct parameter
regimes that they can be studied separately. Through the formation of 
penta-hepta defects, they lead to different types of  transient disordered
states. The results for the dynamics of the  penta-hepta defects shed light
on the persistence of grain  boundaries in such disordered states.

\end{abstract}


\end{frontmatter}


\section{Introduction}
\label{sec:intro}

Roll patterns in Rayleigh-B\'enard convection of a fluid layer heated  from
below have been explored extensively over the years as a paradigmatic
system to study the succession of  transitions from ordered to disordered and
eventually  turbulent states (for a recent review see \cite{BoPe00}).
Substantial  theoretical effort has been devoted to identifying all the linear
instabilities that straight roll patterns can undergo (e.g. \cite{Bu78}).  Depending on the
wavenumber of the roll pattern, the temperature  difference across the layer
(Rayleigh number), and the fluid properties  (Prandtl number) a host of
instabilities has been identified. The main instabilities being the Eckhaus,
zig-zag, cross-roll, oscillatory, and  skew-varicose instabilities. They limit
the band of stable wavenumbers  of straight rolls (Busse balloon). It turns out
that  the oscillatory and the
skew-varicose instabilities depend sensitively on the Prandtl number
and are relevant only if the Prandtl number is sufficiently
small \cite{Bu78}.
 
For small convection amplitudes  a weakly nonlinear description in
terms of a Newell-Whitehead-Segel equation \cite{NeWh69,Se69}
would be expected to be sufficient at least for almost straight roll
patterns. However, it is found that this is true only in the limit of
large Prandtl numbers. It has  been shown that slightly curved roll
patterns drive a mean flow. Because of the
incompressibility of the fluid it induces a non-local coupling of the
rolls, which is not contained in the Newell-Whitehead-Segel
equation \cite{SiZi81}. The strength of the mean  flow increases with
decreasing Prandtl number. For small Prandtl  numbers the
Newell-Whitehead-Segel equation has therefore been extended  to include
an equation for the vertical vorticity characterizing the mean flow
\cite{SiZi81}.
 
Experimentally, the mean flow has been  observed in experiments in a
relatively small circular container \cite{CrLe86}. Since  the usual boundary
conditions require the convection rolls to be  essentially perpendicular to the
side-walls the rolls become strongly  curved in this geometry. The resulting
mean flow has been  identified as the driving force for the
observed persistent creation and annihilation of dislocations
\cite{PoCr85,GrCr88,DaPo89}.

Arguably the most interesting state that is due to the mean flow
associated with small Prandtl numbers is the spiral-defect chaos
observed in large-aspect ratio experiments on thin gas layers
\cite{MoBo93}. It is characterized by the appearance of various types of defects
in the pattern with small rotating spirals being the ones that are visually 
most striking. Spiral-defect chaos does not arise from a linear instability of 
the straight roll pattern and, in fact, bistability of straight roll patterns
and spiral-defect chaos has been observed experimentally
\cite{CaEg97}. The onset of spiral-defect chaos depends strongly on
the Prandtl number, indicating that mean flows play an essential role
in maintaining this state \cite{XiGu93,BeFa93,DePe94a,LiAh96}.                    

Motivated by the strong impact of mean flows on roll convection
patterns we consider in this paper the effect of such flows on the 
stability and dynamics of hexagonal patterns.  Hexagonal patterns are
commonly found in spatially extended non-equilibrium systems such
as non-Boussinesq Rayleigh-B\'{e}nard convection (e.g.
\cite{BoBr91}), Marangoni convection (e.g. \cite{ScVa95}),   Turing
structures in chemical systems \cite{OuSw91a}, crystal growth (e.g.
\cite{CaCa87}), and surface waves on vertically vibrated liquid or
granular layers (Faraday  experiment) \cite{EdFa94,MeUm95}.  Not in
all of these systems mean flows of the  type discussed above arise.
Clearly, they are relevant for the small Prandtl  numbers in gas
convection in very thin layers  \cite{MoBo93}. Interestingly,  the 
skewed-varicose  instability and a (transient) state similar to
spiral-defect chaos have been observed also  in  vertically vibrated  granular
layers \cite{BrBi98,BrLe01}. Since in convection at low Prandtl 
numbers they are a signature of the importance of mean flow, 
mean flow may also be
relevant in vertically vibrated  fluids. We focus in this paper on
patterns arising from a steady bifurcation, as is the case in
convection. Immediately above onset, parametrically excited standing
waves  like those arising in the Faraday experiment behave in many
respects  like patterns that are due to a steady bifurcation (e.g. \cite{Ri90a}).
Our  results may therefore also be relevant for 
hexagon patterns in suitable Faraday experiments.
 
In the absence of mean flows the stability of hexagonal patterns has been 
studied in detail in the weakly nonlinear regime. Starting from  three coupled
Ginzburg-Landau-type equations for the amplitudes of the rolls that make up
the hexagonal pattern two coupled phase equations have been derived that
describe the dynamics of long-wave deformations of the hexagon patterns
\cite{LaMe93,Ho95,EcPe98,Ho00}. In these theoretical analyses two  types of
long-wave instabilities have been identified, a longitudinal and a  transverse
mode, with the longitudinal mode being relevant only for Rayleigh numbers 
very close to the saddle-node  bifurcation at which the hexagons first 
appear. In a more detailed analysis  also perturbations with arbitrary
wavenumber and simulations of the nonlinear evolution ensuing from the
instabilities have been included \cite{SuTs94}. The  instabilities typically lead
to the formation of penta-hepta defects \cite{PiNe93,RaTs94}. Their dynamics
and interaction has also been studied in the absence of mean flows
\cite{Ts95,Ts96}. In the strongly nonlinear regime long-wave perturbations
are still captured by phase equations \cite{Ho00}.  Specifically for
convection at large Prandtl numbers driven by a combination of buoyancy 
and surface tension the stability of hexagons and their dynamics has  been
investigated in \cite{Be93}. Experimentally, the stability of hexagonal convection
patterns has not been  studied in detail. A contributing factor has been that
controlled changes in the wavenumber of the pattern are considerably more
difficult to effect than in  systems like Taylor-vortex flow, where detailed agreement
between experiment and theory has been achieved (e.g.
\cite{DoCa86,RiPa86}). Recently, however, it has been possible to use
localized heating as a  printing technique for convection patterns
\cite{Scpriv,Bopriv}. This  will make detailed analyses of the
wavenumber-dependent instabilities  of hexagon patterns experimentally
accessible.
   
\begin{figure}
\centerline{\epsfxsize=06.0cm\epsfysize=06.5cm\epsfbox{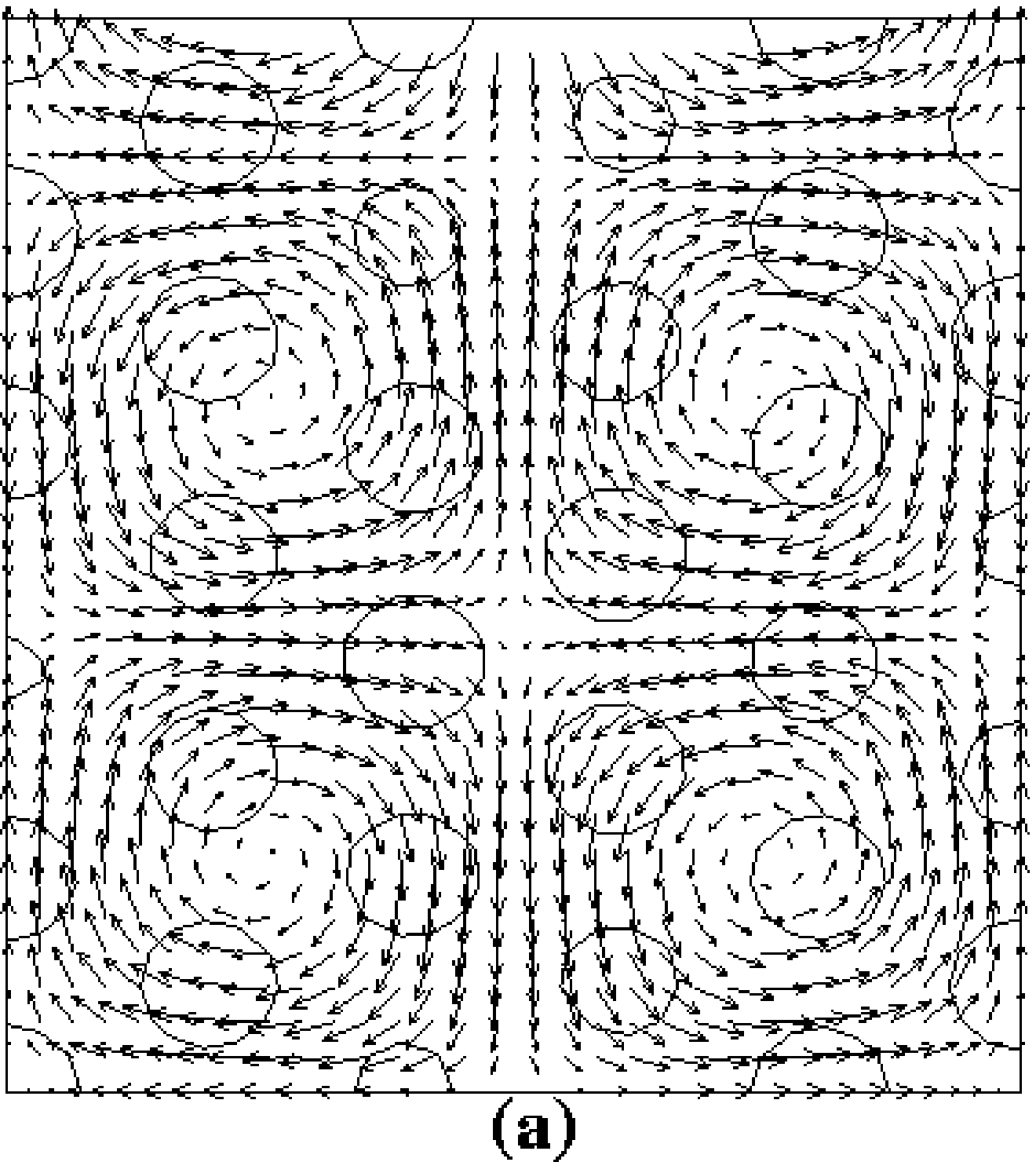}
\hspace{1cm}\epsfxsize=06.0cm\epsfysize=06.5cm\epsfbox{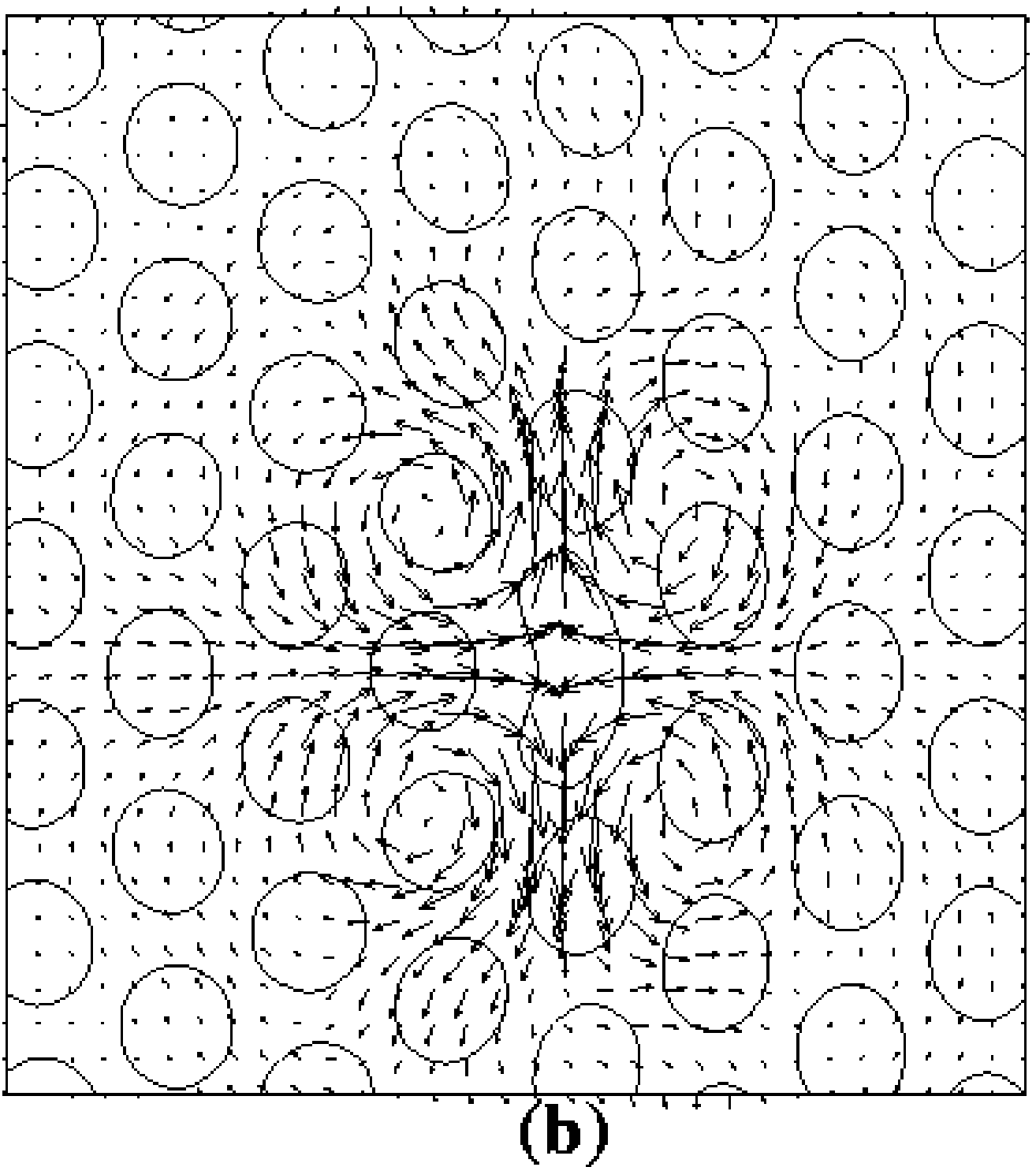}}
\caption{
Illustration of the mean flow due to modulation of the hexagon envelope
(panel(a)) or due to a penta-hepta defect (panel(b)).  For visual clarity the 
wavelength of the hexagon pattern has been chosen larger (relative to the
modulation wavelength) than is appropriate for the weakly  nonlinear theory
presented in this paper.  
}
\label{fig:mean_ill}
\end{figure}       
 
The goal of this study is to investigate the  linear stability of hexagons
coupled to a two-dimensional mean flow driven by modulations of the
pattern and the nonlinear evolution arising from the  instabilities.  An
example of the flow generated by two typical patterns is shown in
fig.\ref{fig:mean_ill}a,b. In fig.\ref{fig:mean_ill}a a periodic hexagon
pattern is sinusoidally modulated on a  length scale five times the
hexagon size and fig.\ref{fig:mean_ill}b shows the mean flow
generated by a penta-hepta defect. The mean flow is found to couple
only to one of the two long-wave  modes. As a  consequence,  for
sufficiently small Prandtl numbers each of the  two long-wave
instabilities dominates in a separate, experimentally accessible
parameter regime. Both instabilities  induce the formation of defects.
We study their motion and briefly  touch upon its impact on
disordered patterns and grain boundaries. 

The paper is organized as follows: first we formulate the problem extending
previous work on the mean flow generated by roll convection
\cite{DePe94,Be94}. We then study the stability of the hexagonal pattern with
respect to long-wave perturbations (sec.\ref{sec:linear01}) and with  respect
to general side-band perturbations (sec.\ref{sec:linear02}). In section
\ref{sec:num}   the nonlinear evolution of the instabilities including the
formation, dynamics, and annihilation  of defects is investigated.
There, we also study the transition from hexagonal to roll  patterns
triggered by the side-band instabilities.                                     

\section{Amplitude equations}
\label{sec:amp}
The interaction between weakly nonlinear convection rolls and the
mean flow generated by them has been investigated for
Rayleigh-B\'{e}nard convection with both no-slip and stress-free
boundary conditions at the top and bottom plate
\cite{DePe94,Be94}. The relevant amplitude equations were derived and used 
to determine the  stability of rolls with respect to side-band instabilities.
Expanding the fluid velocity in terms of the complex amplitude $A$ of the convection rolls
 and the real streamfunction $Q$ for the two-dimensional mean flow as 
($\epsilon \ll 1$),
\bea
{\bf v}(\tilde{x},\tilde{y},\tilde{z},\tilde{t})=\epsilon A(x,y,t) 
e^{iq_c\tilde{x}}{\bf v}_q(\tilde{z})+\epsilon^2\left\{ 
\vec{\nabla}\times(-Q\hat{z})f(\tilde{z})\right\}+h.o.t.,
\label{eq:vexpand}
\eea  
the extended Ginzburg-Landau equation for $A$ and $Q$ is given by
\begin{eqnarray}
\label{eq:amp_roll1}
\partial_t A &=&
(\mu   +(\partial_x - i \lambda\partial^2_y)^2  -|A|^2)A- 
i s_1 A\partial_yQ + h.o.t. \\
\label{eq:amp_roll2}
M(Q) &=&  2q_1(\partial_x-i\lambda\partial^2_y)\partial_y |A|^2 + h.o.t.
\end{eqnarray}
where $\lambda\equiv \epsilon/2q_c$ with $\epsilon$ the supercriticality parameter 
and $q_c$ the critical wavenumber.
Here $M(Q)=\nabla^2 Q$ for no-slip boundary conditions \cite{DePe94}
and $M(Q)=(q_2\nabla^2\partial_t -\nabla^4)Q$  for the stress-free case
\cite{Be94}. It is worth noting that in the no-slip case the mean flow is not an
additional dynamical variable and could  in principle be adiabatically
eliminated which would lead to a non-local equation for $A$. The
slow time $t$ is given by $t=\epsilon^2 \tilde{t}$. The non-isotropic
Newell-Whitehead-Segel  operator $(\partial_x - i
\lambda\partial^2_y)^2$ reflects the anisotropic scaling of the slow
spatial scales $x=\epsilon \tilde{x}$ and  $y=\epsilon^{1/2}\tilde{y}$.
To leading order, the mean flow is driven by variations in the
magnitude of the convective amplitude and couples back through the 
advective term $i s_1 A\partial_yQ=is_1 A V_x$, where $V_x$ is the 
$x$-component of the mean flow. At higher orders a term of the form 
$V\cdot \nabla A$ would arise as well.

For the description of hexagonal convection patterns coupled to the  mean 
flow the treatment has to be extended to include rolls in other directions than the
$x$-axis.  Using a coordinate transformation, it is straightforward to
generalize the coupling term $A\partial_y Q$ for rolls in the $x$-direction $\hat
n_1\equiv (1,0)$ to rolls in the other two hexagonal directions  $\hat n_2 \equiv (-1/2,
\sqrt{3}/2)$ and  $\hat n_3\equiv(-1/2,-\sqrt{3}/2)$.   The amplitudes $A_i$,
$i=1,2,3$, correspond then to rolls with wave vector 
$q_c\hat{n}_i$.  For the generalization of the advection term $iA\partial_y Q$ in
(\ref{eq:amp_roll1}) it is convenient to introduce the unit vector $\hat{\tau}_i$
perpendicular to $\hat{n}_i$. To complete the extension of 
eqs.(\ref{eq:amp_roll1},\ref{eq:amp_roll2}) from roll to hexagonal patterns
 we also add non-Boussinesq quadratic terms. For no-slip boundary conditions, 
the minimal set of equations describing hexagons coupled to a mean flow then reads
\begin{eqnarray}
\label{eq:mf1}
\partial_t A_j &=&
\left(\mu+ (\hat{n}_j\cdot\nabla)^2 - 
|A_i|^2-\nu (|A_{j-1}|^2+|A_{j+1}|^2)\right) A_j+ \nonumber\\
&& A_{j-1}^*A_{j+1}^* - i \beta A_j (\hat \tau_j\cdot \nabla) Q,\\ 
\label{eq:mf2}
\nabla^2 Q &=& -\left\{-2\partial_{xy} |A_1|^2 +\right.  \nonumber \\
     && 
 \left. \left[-\frac{\sqrt{3}}{2}(-\partial_y^2+\partial_x^2)+\partial_{xy}\right] |A_2|^2 +
  \left[ \frac{\sqrt{3}}{2}(-\partial_y^2+\partial_x^2)+\partial_{xy}\right] |A_3|^2 
\right\},
\end{eqnarray}
with cyclic permutation on $j$. It is not clear how the different  scaling
of the longitudinal spatial variable in the direction  of $\hat{n}_i$ and
of the transverse variable in the direction of  $\hat{\tau}_i$ can be
implemented in the hexagonal case where these  directions are
different for each of the three hexagon modes. We  therefore use an
isotropic scaling, $x=\epsilon\tilde{x}$ and  $y=\epsilon\tilde{y}$,
which renders the derivatives $i\lambda(\hat\tau_j\cdot\nabla)^2$
higher order in $\epsilon$. This can lead to degeneracies in  certain
growth rates like that of the zig-zag instability  (e.g.
\cite{CaKn01,EcRi01}). Then higher-order terms have to be
re-introduced. For the hexagonal patterns this is,  however, not the
case. Note that while there is no direct damping  of transverse
variations, each of the $A_j$ is still allowed to  vary in its respective
transverse direction. This variation induces a  variation in the
longitudinal direction of the other two modes, in  particular through 
the resonant interaction term $A_{j-1}^*A_{j+1}^*$, which then
induces damping. 

The cubic coefficient $\nu$ decreases monotonically  with Prandtl
number from $\nu\sim 2.0$ for $\Pr=0.5$ to $\nu\sim 1.5$ for $ \Pr
\sim 10$  \cite{Bu67,BoBr91}. For the numerical results in this paper
we fix $\nu=2$. Note that the mean flow amplitude $Q$  is of higher
order than the hexagonal amplitudes, in fact, $Q\propto {\cal
O}(A^2)$.  Nevertheless, the coupling term
$iA(\hat{\tau}_j\cdot\nabla)Q$ is retained because $|\beta|$ can be
much larger than one for low Prandtl number.  For example,
$\beta\sim -18 $ for Prandtl number $\sim 0.1$ \cite{DePe94}. 
Furthermore, the mean flow introduces qualitatively new effects since 
it provides a non-local coupling between the convection amplitudes
that  is due to the incompressibility of the fluid. 


In the absence of the mean flow,  there exists a Lyapunov functional
${\cal F}$ for (\ref{eq:mf1}), which guarantees that only stationary
patterns  are possible as $t\rightarrow \infty$. Simple stationary
solutions are:
\begin{eqnarray}
\label{eq:ss1}
&&{\rm (i)\,\,   rolls }
\;\;\;A_j = (\mu - q^2)^{1/2},\;\; A_{j\pm1}=0, \\
\label{eq:ss2}
&&{\rm (ii)\,\,  hexagons}
\;\;\;A_1=A_2=A_3=\frac{1\pm \sqrt{1+4(\mu-q^2)(1+2\nu)}}{2(1+2\nu)},\\
\label{eq:ss3}
&&{\rm (iii)\,\, mixed\,\, modes}
\;\;\;A_j=\frac{1}{\nu-1},\;\; 
A_{j\pm1}=\sqrt{\frac{\mu-q^2-A_1^2}{1+\nu}}.
\end{eqnarray}
As usual, the hexagon solution  given by equation (\ref{eq:ss2}) with a minus
sign in  front of the square root is not stable.  Thus, in the 
following we focus on the hexagon solution with the plus sign.

Since the mean flow couples to hexagons only through a modulation in the
amplitudes, the  stationary states (\ref{eq:ss1}-\ref{eq:ss3}) with constant
amplitudes  are also solutions to equations (\ref{eq:mf1},\ref{eq:mf2}).
However, their stability is modified by the coupling to the mean flow.  In the
case of rolls it can lead to an  oscillatory instability (oscillatory skew-varicose
instability), implying that there is no Lyapunov  functional when
taking the mean flow into account.   


\section{Linear Stability Analysis: Long-wave Approximation}
\label{sec:linear01}

We conduct a linear stability analysis of the hexagonal pattern (\ref{eq:ss2})
in the long-wave limit  following the procedures in \cite{Ho95}. The
amplitudes are perturbed around a hexagon pattern,
\begin{equation}
A_j=R_0 e^{i q \hat{n}_j\cdot(x,y)} (1 + r_j + i \phi_j),\;\;
j=1,\;2,\; 3,\label{eq:pert}
\end{equation}
where $R_0\equiv \left(1+\sqrt{1+4(\mu-q^2)(1+2\nu)}\right)/2(1+2\nu)$ 
is the amplitude of the
stationary solution with reduced wavenumber $q$,
and $r_j$ and $\phi_j$ are the amplitude and phase perturbations, respectively.
We also define $u$ and $v$ as in \cite{Ho95}
\begin{equation}
\label{eq:u_v}
u=R_0^2(1-\nu)+R_0,\;\;\;
v=2R_0^2(1+2\nu)-R_0,
\end{equation}
where $v=0$ corresponds to
the saddle-node bifurcation at which the hexagons come into existence. 
It is equivalent to 
\bea
\mu_{SN}=q^2-\frac{1}{4}\frac{1}{1+2\nu}.\label{eq:sn}
\eea
At $u=0$ the hexagons become unstable to the mixed mode solution
(\ref{eq:ss3}).  It is equivalent to 
\bea
\label{mm_solution}
\mu_{MM}=q^2+\frac{\nu+2}{(\nu-1)^2}
\eea 
Both $u$ and $v$ have to be greater than zero for
hexagons to be stable.
 
We first rewrite (\ref{eq:mf2}) in terms of the perturbation amplitudes.
The mean flow then satisfies the linearized equation 
\begin{eqnarray}
\label{eq:lq}
\nabla^2 Q &=& -2R_0^2\{ -2\partial_{xy}r_1 +	\nonumber\\
   && \left[ \frac{-\sqrt{3}}{2}
       (-\partial^2_y+\partial^2_x)+\partial_{xy} \right]r_2 +
      \left[ \frac{ \sqrt{3}}{2}
       (-\partial^2_y+\partial^2_x)+\partial_{xy} \right]r_3 \}.
\end{eqnarray}
Substituting the perturbed amplitudes  (\ref{eq:pert}) into
(\ref{eq:mf1}),  we obtain
\begin{eqnarray}
\label{eq:ll1}
\partial_Tr_{1} &=& 
     -\Lambda r_1 +
      \Lambda_1(r_2+r_3)+(\ninabla)^2 r_1 -2q(\ninabla)\phi_1,\\
\label{eq:ll2}
\partial_Tr_{2} &=& 
     -\Lambda r_2 + 
      \Lambda_1(r_3+r_1)+(\njnabla)^2 r_2 -2q(\njnabla) \phi_2,\\
\label{eq:ll3}
\partial_Tr_{3} &=& 
     -\Lambda r_3 + 
      \Lambda_1(r_1+r_2)+(\nknabla)^2 r_3 -2q(\nknabla)\phi_3,\\
\label{eq:ll4}
\partial_T\phi_{1}&=&-R_0\Phi + 
2q(\ninabla) r_1 + (\ninabla)^2\phi_1 - s_1 \partial_y Q,\\
\label{eq:ll5}
\partial_T\phi_{2}&=&-R_0\Phi + 
2q(\njnabla) r_2 + (\njnabla)^2\phi_2 -
s_1\left(\frac{-\sqrt{3}}{2}\partial_x-\frac{1}{2}\partial_y\right)Q,
\\
\label{eq:ll6}
\partial_T\phi_{3}&=&-R_0\Phi + 
2q(\nknabla) r_3 + (\nknabla)^2\phi_3 -
s_1\left(\frac{\sqrt{3}}{2}\partial_x-\frac{1}{2}\partial_y\right)Q,
\end{eqnarray}
where $\Lambda=(1+2R_0)R_0$, $\Lambda_1=(1-2\nu R_0)R_0$.

In this section we focus on the limit of long-wave perturbations and
introduce super-slow scales $X=\delta x$, $Y=\delta y$, and 
$T=\delta^2t$ and a slow gradient, $\nabla=\delta\tilde{\nabla}$
($\delta \ll  1$).  In  this limit the amplitude perturbations $r_j$ and the
global phase $\Phi\equiv\phi_1+\phi_2+\phi_3$  are slaved to the
two  slow modes $\phi_x\equiv -(\phi_2+\phi_3)$ and  $\phi_y\equiv
(\phi_2-\phi_3)/\sqrt{3}$, which arise from the translation  symmetry
of the system. Adiabatic elimination reduces then
(\ref{eq:ll1}-\ref{eq:ll6}) to
\begin{eqnarray}
\label{eq:slave1}
r_1 &=& r_3 - \frac{q}{u}((\niig)\phi_1-(\nkkg)\phi_3),\\
r_2 &=& r_3 - \frac{q}{u}((\njjg)\phi_2-(\nkkg)\phi_3),\\
r_3 &=& -\frac{2q}{v}(\nkkg)\phi_3 -
\frac{q(2u-v)}{3uv}\left(
\sum_{i=1}^3(\hat{n}_i\cdot\hat\nabla)\phi_i-3(\nkkg)\phi_3\right),\\
\label{eq:slave2}
3R_0\Phi &=& \sum_{i=1}^3(\hat{n}_i\cdot\hat\nabla)^2\phi_{i} 
+2q\sum_{i=1}^3 (\hat{n}_i\hat\nabla)r_i,
\end{eqnarray}
and an evolution equation for the phase vector
$\vec{\phi}=(\phi_x,\phi_y)$, 
\begin{equation}
\label{eq:phase}
\partial_t\vec{\phi} = D_{\perp} \hat{\nabla}^2 \vec{\phi} + 
(D_{\|}-D_{\bot}) \hat{\nabla}(\hat{\nabla}\cdot \vec{\phi}) +
\frac{3q\beta 
R_0^2}{2u}\hat{\nabla}\times(\hat{\nabla}\times\vec{\phi}).
\end{equation}
The last term of the phase equation (\ref{eq:phase}) arises from the 
coupling to the mean flow, which is driven by the curl of the phase,
\begin{equation}
\label{meanQ_lw}
 Q = -\frac{3 R_0^2 q}{2u}(-\hat{e}_z\cdot\hat{\nabla}\times\vec{\phi}).
\end{equation}
In (\ref{eq:phase}) $D_{\|}$ and $D_{\bot}$ are the longitudinal and 
transverse phase diffusion coefficients for the hexagons in the absence 
of a mean flow \cite{LaMe93,EcPe98,EcRi00}.
 
Even in the presence of the mean flow, equation (\ref{eq:phase}) allows a
decomposition of the phase vector  $\vec{\phi}$   into  a longitudinal
(curl-free) mode $\vec{\phi}_l$ satisfying
\bea
\label{eq:curl_free}
\hat{e}_z\cdot\hat{\nabla}\times\vec{\phi_l}=0,
\eea
which has growth rate 
\bea
\sigma_l =\frac{{\sf k}^2}{2}
\left[ -\frac{3}{2}+\frac{q^2}{u}+\frac{4q^2}{v}\right],
\label{eq:long}
\eea
and a transverse (divergence-free) mode $\vec{\phi}_t$ satisfying
\bea
\label{eq:div_free}
\hat{\nabla}\cdot \vec{\phi}_t =0
\eea
with growth rate
\bea
\label{eq:transverse}
\sigma_t= \frac{{\sf k}^2}{2}
\left[ -\frac{1}{2}+\frac{q^2}{u} -\frac{3q\beta R^2_0 }{u}\right].
\eea
To this order the growth rates do not depend on the orientation of the 
perturbation wave vector $\vec{\sf k}\equiv (k,l)$ but only on its magnitude ${\sf 
k}=\sqrt{k^2+l^2}$.

\begin{figure}
\centerline{\epsfxsize=07.0cm\epsfysize=06.5cm\epsfbox{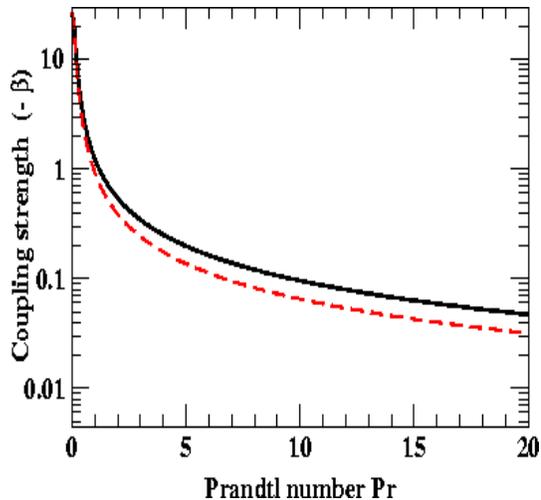}}
\caption{
The coupling strength $\beta$ for rolls. Solid line from \cite{DePe94}. Dashed
line is the dependence conjectured in \cite{ZiSi82} with the free constant 
chosen to match the solid line at $\Pr=0$.  }
\label{fig:beta_pr}
\end{figure}

As expected from equation (\ref{meanQ_lw}), the mean flow only affects 
the transverse mode. It shifts the associated stability boundary of
the hexagons and makes it asymmetric in $q$. The sign of $\beta$ determines 
whether  the
mean flow destabilizes the transverse mode for positive or for
negative  wavenumbers.  For roll convection with no-slip boundary
conditions $\beta$ is always negative \cite{DePe94}  and its
dependence on the Prandtl number is shown in figure \ref{fig:beta_pr}.
We note that $\beta\sim -1$ for gases (unity Prandtl numbers),  and
$\beta\sim -0.1$ for water (Prandtl number $\sim 10$).

\begin{figure}
\centerline{\epsfxsize=07.0cm\epsfysize=07.0cm\epsfbox{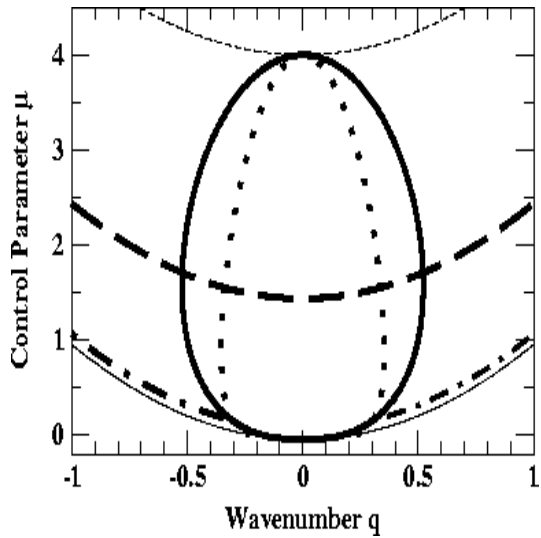}}
\caption{
Summary of the stability boundaries for infinite Prandtl numbers  ($\beta=0$)
showing the saddle-node bifurcation (thin solid line), transition of hexagons
to mixed-mode solution (thin dashed), longitudinal mode (thick solid),
transverse mode (thick dotted), equal-energy line for hexagons  and rolls
(thick dashed), cross-over from longitudinal to transverse mode (thick
dashed-dotted). 
}
\label{fig:no_mf_sb}
\end{figure}

The stability boundaries for infinite Prandtl number (no mean flow)  are
reviewed in figure \ref{fig:no_mf_sb}. Hexagons exist  above the thin solid
line ($v=0$).  The thick solid line and the dotted line are the stability
boundaries for the longitudinal and transverse modes, respectively. Along
the thick dashed line, rolls and hexagons have the same value of the
Lyapunov functional. Hexagons become unstable to rolls ($via$ the  unstable
mixed mode (\ref{eq:ss3})) at the thin dashed line ($u=0$). Note that  since we
have scaled the coefficient of the quadratic nonlinearity in 
(\ref{eq:mf1},\ref{eq:mf2}) to 1, this transition occurs at $\mu=4$.  For weak
non-Boussinesq effects this corresponds to very small  convection
amplitudes. As discussed below (see also sec.\ref{sec:linear02}), the decomposition into transverse and
longitudinal  mode does not hold beyond the long-wave limit. It is replaced
by a  separation into a `wide-splitting' and a `narrow-splitting' mode 
\cite{SuTs94}. The dashed-dotted line separates the regions in which the 
wide-splitting and the narrow-splitting mode represents the mode with 
maximal growth rate, respectively. In view of the effect of the  mean
flow it is worth noting that the regime in which the longitudinal 
mode is the relevant destabilizing mode is typically extremely  small
(below  dashed-dotted line in fig.\ref{fig:no_mf_sb}) and it is very difficult to
investigate that instability  experimentally. In fact, even in numerical
simulations of (\ref{eq:mf1}) (with $\beta=0$) we found it difficult to 
separate the dynamics of the two modes (see also \cite{SuTs94}). Note that  due to
the scaling of the quadratic coefficient in  (\ref{eq:mf1},\ref{eq:mf2}) the
usually small range over which hexagons  can be observed extends from
$\mu \approx 0$ to $\mu \approx 1.5$. 
      
\begin{figure}
\centerline{\epsfxsize=7.0cm\epsfysize=7.0cm\epsfbox{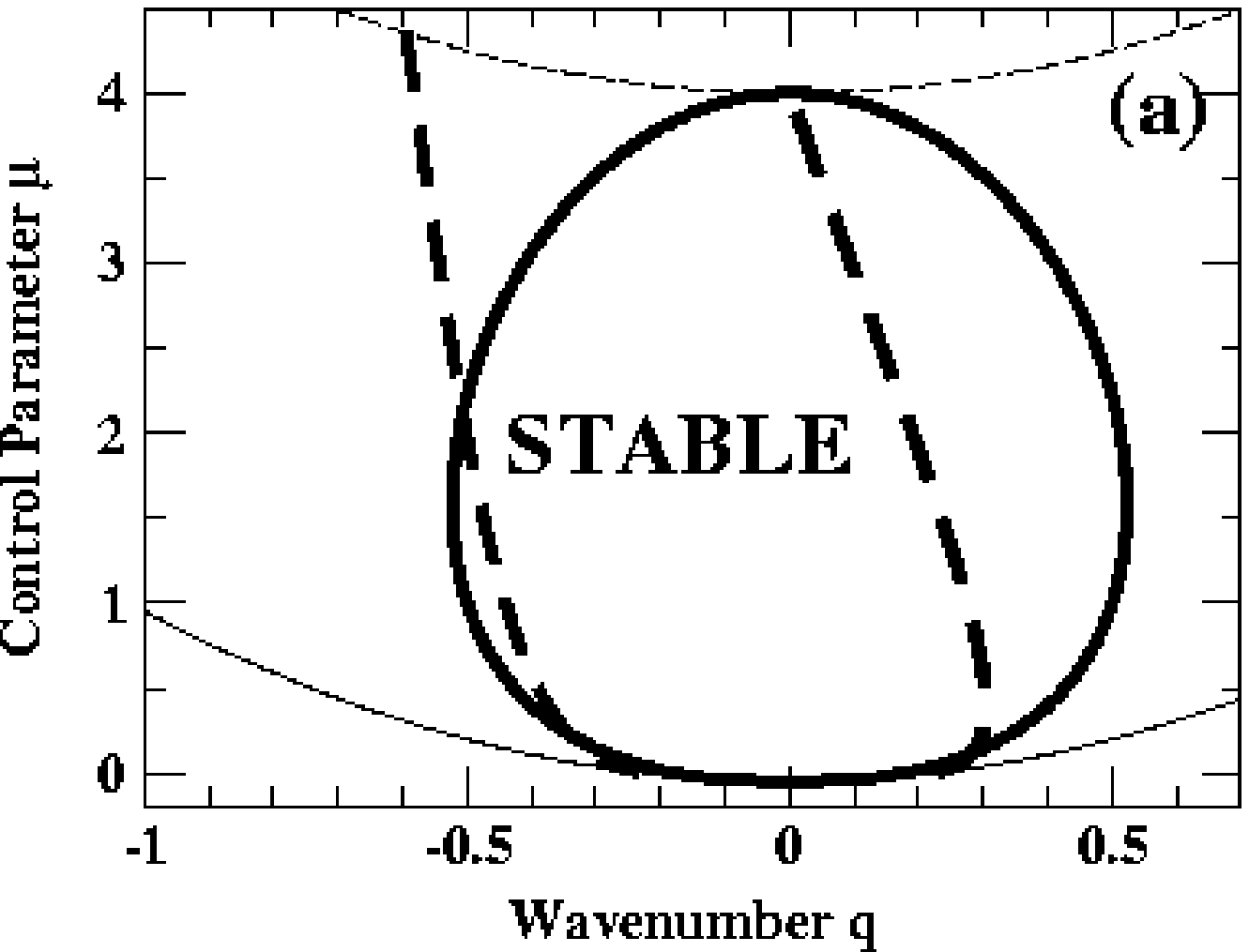}
\hspace{1cm}\epsfxsize=7.0cm\epsfysize=7.0cm\epsfbox{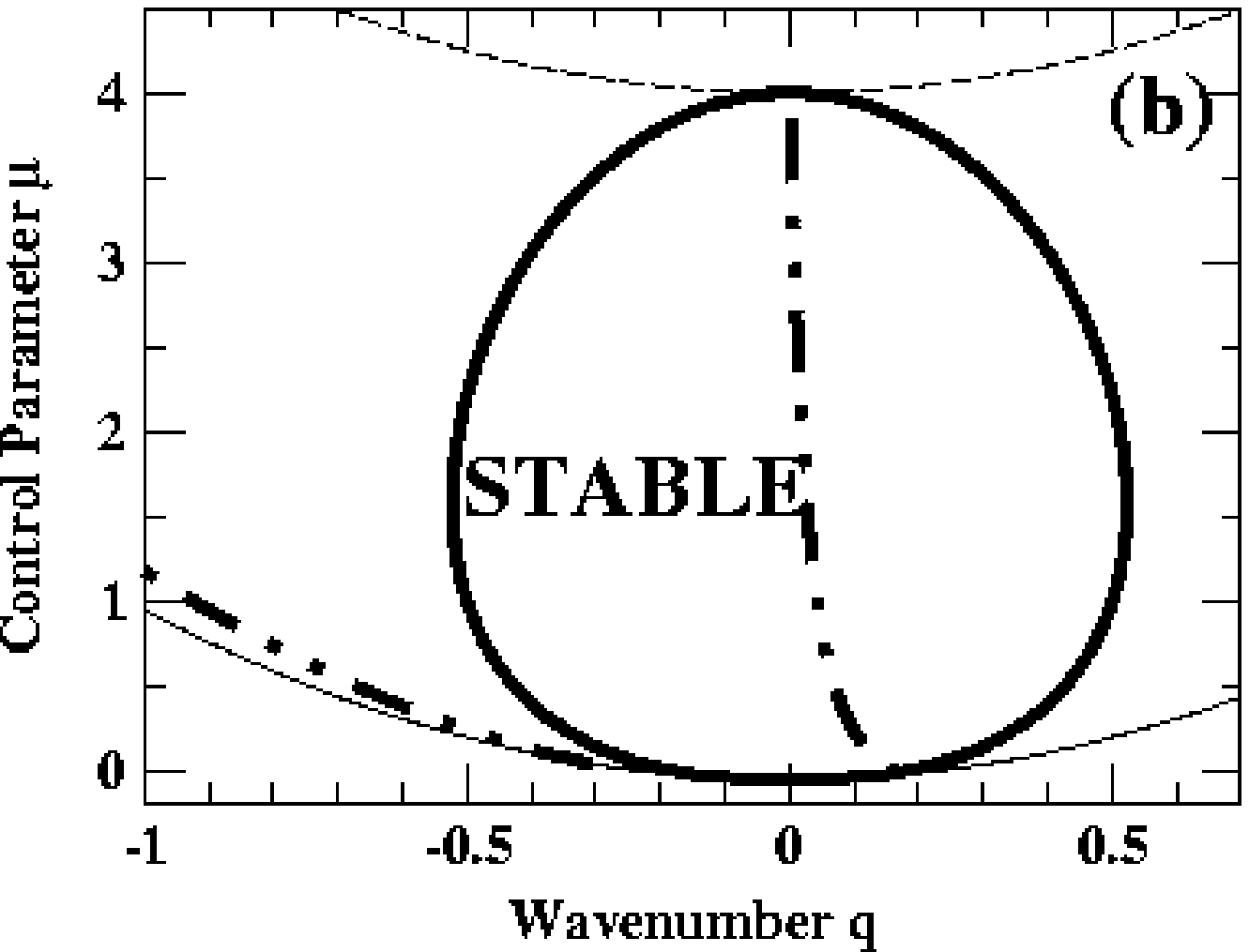}}
\centerline{
            \epsfxsize=7.0cm\epsfysize=5.5cm\epsfbox{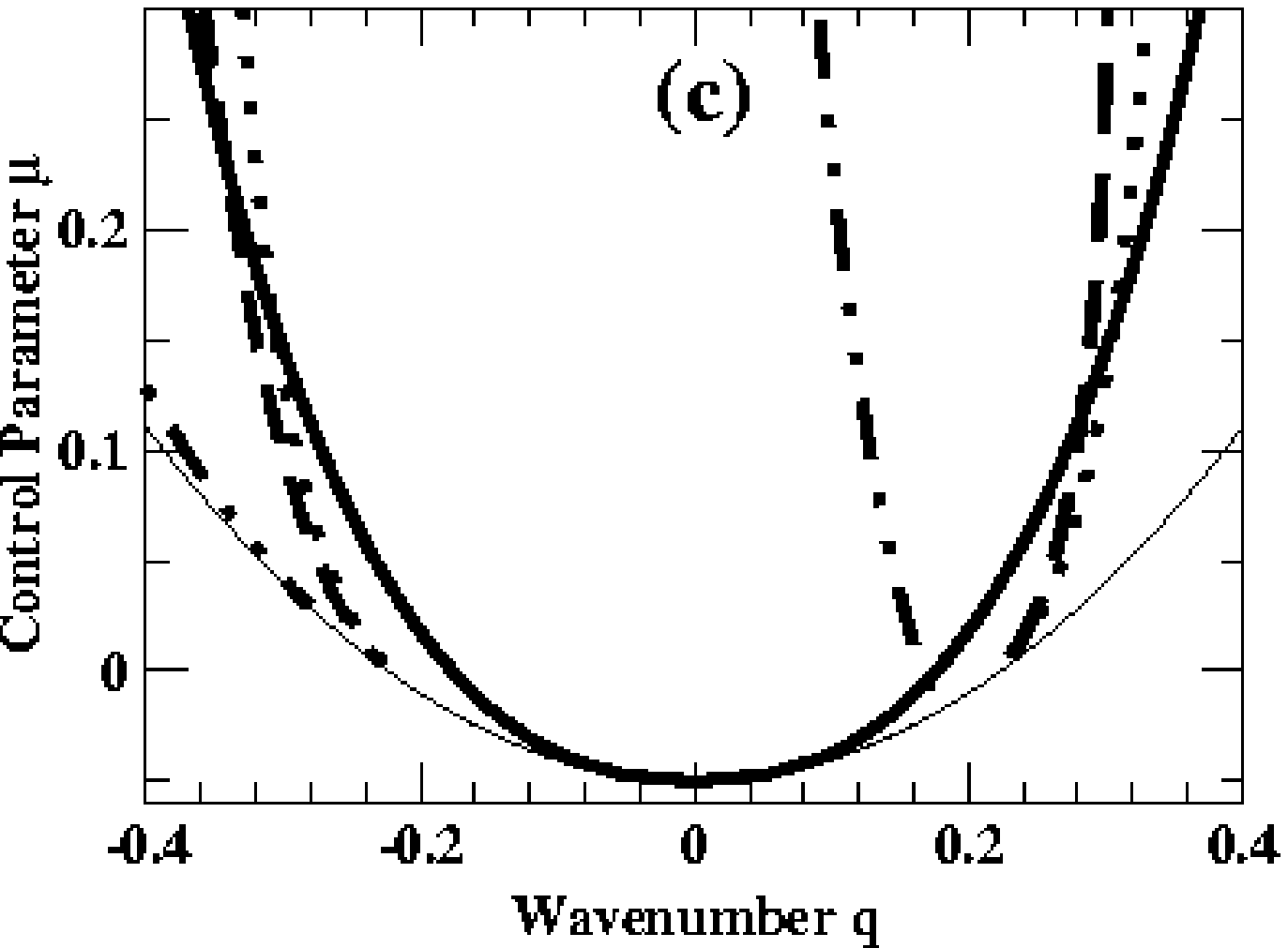}}
\caption{Stability boundaries of hexagons coupled to a two-dimensional
mean flow. The thick solid line is the stability limit given by the longitudinal 
mode. The thick broken lines are for the transverse mode. Thin lines are as  in
fig.\protect{\ref{fig:no_mf_sb}}. Panel (a) is for $\beta=-0.2$ ($\Pr\sim 5$) and panel
(b) is for $\beta=-3$ ($\Pr\sim 0.5$). Panel(c): enlargement of the
stability boundaries for small $\mu$  showing results for $\beta=0$,
$\beta=-0.2$, and $\beta=-3$. 
}
\label{fig:sb_meanfl_lw}
\end{figure}  

Figure \ref{fig:sb_meanfl_lw}a,b shows how the stability boundaries are
altered by the mean flow for $\beta=-0.2$ and $\beta=-3.0$, respectively.
Figure \ref{fig:sb_meanfl_lw}(c) shows an enlargement of figure
\ref{fig:sb_meanfl_lw}a,b for small $\mu$. As discussed before, the mean flow
affects only the transverse mode. The corresponding stability limits  are
denoted by thick dotted, dashed and dashed-dotted lines for  $\beta=0$,
$\beta=-0.2$, and $\beta=-3.0$, respectively. The thick  solid line gives the
stability limit due to the longitudinal mode,  which is independent of $\beta$.
For all values of $\beta$ it is the  relevant instability immediately above the
saddle-node bifurcation at  which the hexagons first arise 
(cf. fig.\ref{fig:sb_meanfl_lw}c). For  somewhat
larger values of $\mu$ the transverse mode becomes relevant as well. In the
physically relevant case, $\beta<0$, it is destabilized by the mean  flow for
positive $q$ strongly reducing the stability balloon there. While for  large
Prandtl numbers the cross-over from the longitudinal to the  transverse mode
is almost imperceptible in the stability limit itself,  it occurs very suddenly
when the mean flow is important. This behavior  has also been found in fully
nonlinear stability calculations of  hexagonal convection in non-Boussinesq
convection \cite{MoPe02}. For $q<0$ the transverse mode is stabilized by
the mean flow. This allows the longitudinal mode  to become important again
for larger $\mu$. In fact, for sufficiently  large $\beta$ the transverse mode
does not contribute to the stability  limits at low $q$ for any value of $\mu$
(cf. dashed-dotted line for $\beta=-3$ in fig.\ref{fig:sb_meanfl_lw}b). 
Thus, for small Prandtl numbers the
mean flow makes it possible to  investigate the two instabilities separately. In
sec.\ref{sec:num01} below we discuss the different nonlinear behavior arising
from the two instabilities.

The leading-order phase equation (\ref{eq:phase}) is isotropic. Only at
higher order in the gradients this symmetry  is reduced to that of the
underlying hexagon pattern. Similarly, at  higher orders (i.e. for larger
magnitudes of the perturbation  wavevector) the eigenmodes of the phase
equation are not expected to  be purely transverse or longitudinal any more.
In sec.\ref{sec:linear02} we determine the character of the destabilizing
modes  numerically  for finite perturbation  wavevectors, but only for specific
values of the parameters. To obtain  more general results for the
influence of the mean flow  on the destabilizing modes we derive here the phase
equation to fifth order. 

If only one of the two modes is unstable the coupled
phase equations can  be reduced to a single equation for the unstable mode
\cite{Ho95}. It is obtained  by eliminating the stable mode adiabatically. Since
both modes are  long-wave modes satisfying diffusion-like equations the
elimination  introduces non-local terms in the equation for the unstable mode.
To demonstrate more explicitly the breaking of the isotropy and the mixing  of the two modes
we derive the coupled fifth-order equations  for both
components of the phase vector $\vec{\phi}$. This can be done 
systematically in the vicinity of the codimension-two point $(q^{(ct)}, \mu^{(ct)})$
where both $\sigma_l$ and $\sigma_t$ are zero to cubic order  (cf.
(\ref{eq:long},\ref{eq:transverse})). There the orientation of the  eigenmodes
depends solely on the fifth-order terms. Away from the  codimension-two
point it is determined by the competition between  the cubic and the
fifth-order terms.

To facilitate the analysis we restrict to cases of 
weak mean flow and expand in $|\beta|\ll 1$.  To cubic order, the 
codimension-two point is given by  $q^{(ct)}=q_0+\beta q_1$ and
$R_0^{(ct)}=R_{00}+\beta R_{01}$ with 
\begin{eqnarray}
\label{eq:codim2_beta}
q_0=\frac{\sqrt{2\nu+2}}{4\nu},\;\; q_1=\frac{3(\nu+1)}{8\nu^3}\;\;\mbox{and}\;\;
R_{00}=\frac{1}{2\nu}, \;\;   R_{01}=\frac{3\sqrt{2\nu+2}}{8\nu^3}.
\end{eqnarray}
Here we consider the magnitude $R_0$ as the control parameter instead of
$\mu$. To first order in $\beta$,  the fifth-order, linear phase equations at the
codimension-two point read as follows:
\begin{eqnarray}
\label{eq:phase:1}
\partial_t\nabla^2\phi_x&& = -\frac{\nu^2}{8(\nu+1)}\nabla^2\left[(4\partial^3_x\partial_y+12\partial_x\partial^3_y)\phi_y
+(11\partial^4_x+3\partial^4_y+6\partial^2_x\partial^2_y)\phi_x\right]\nonumber \\
&&+\beta\frac{\sqrt{2(\nu+1)}}{16(\nu+1)^2}\left[\left(3(4+13\nu)\partial^5_x\partial_y+6(8+3\nu)\partial^3_x\partial^3_y+
                                             9(4+3\nu)\partial_x\partial^5_y\right)\phi_y \right.\nonumber\\
&&          \left.                       +\left(3(11+7\nu)\partial^6_x+3(17+26\nu)\partial^4_x\partial^2_y+
                                             9(3+\nu)\partial^2_x\partial^4_y+9\partial^6_y\right)\phi_x\right],
\end{eqnarray}
\begin{eqnarray}
\label{eq:phase:2}
\partial_t\nabla^2\phi_y&& = -\frac{\nu^2}{8(\nu+1)}\nabla^2
\left[(\partial^4_x+9\partial^4_y+18\partial^2_x\partial^2_y)\phi_y
+(4\partial^3_x\partial_y+12\partial_x\partial^3_y)\phi_x\right]\nonumber \\
&&+\beta\frac{\sqrt{2(\nu+1)}}{16(\nu+1)^2}\left[\left(3\partial^6_x+3(19+13\nu)\partial^4_x\partial^2_y+
 9(9+2\nu)\partial^2_x\partial^4_y+27\partial^6_y\right)\phi_y \right. \nonumber\\
&&\left. +\left(3(4+7\nu)\partial^5_x\partial_y+3(8+13\nu)\partial^3_x\partial^3_y+
                                             9(4+\nu)\partial_x\partial^5_y\right)\phi_x\right].
\end{eqnarray}              
A brief summary of the derivation and the full nonlinear phase equations  (for
$\beta=0$) are included in the appendix. Similar to the nonlinear phase equations
derived in \cite{Ho95}, eqs.(\ref{eq:phase:1},\ref{eq:phase:2}) are non-local.
Here, however, the reason for this is the nonlocal effect of the mean 
flow: for $\beta=0$ the equations become local. 

\begin{figure}
\centerline{\epsfxsize=4.5cm\epsfysize=5.5cm\epsfbox{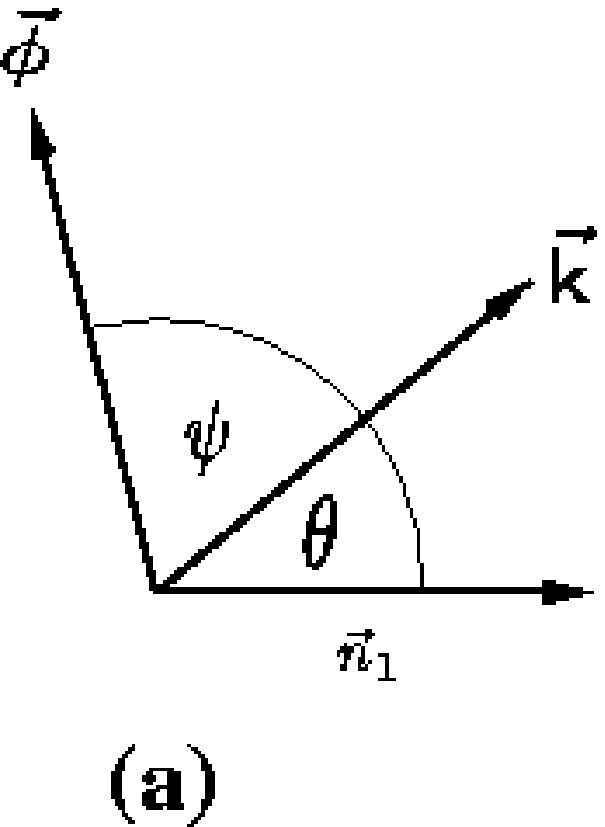}
\hspace{1cm}\epsfxsize=8.5cm\epsfysize=5.5cm\epsfbox{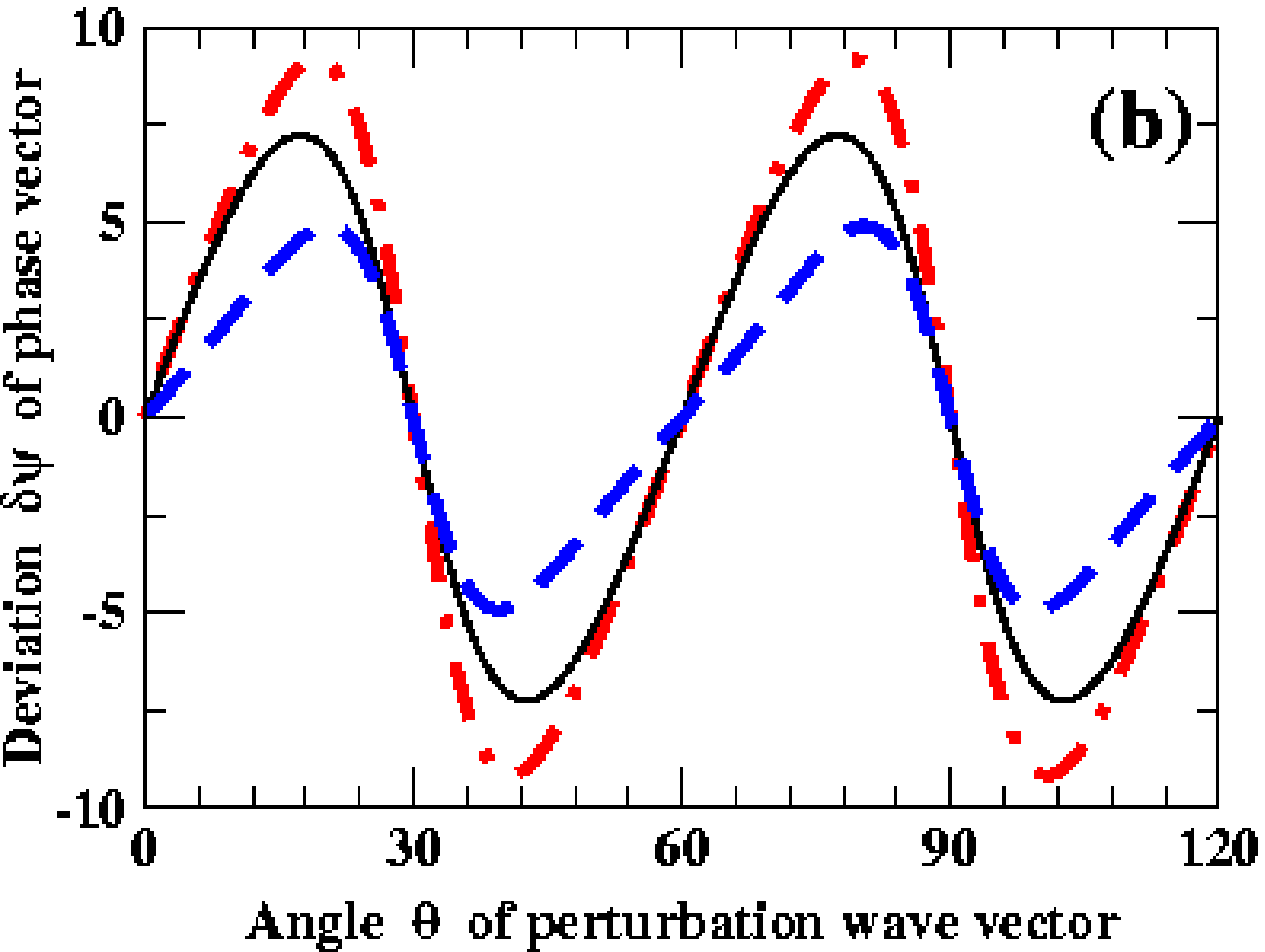}}
\caption{Panel (a): angles $\psi$ and $\theta$. 
Panel (b): Angle deviation $\delta\psi$ for the two eigenvectors at the 
codimension-two point as a function of the angle $\theta$ between $\vec{\sf
k}$ and $\vec{n}_1$ for $\nu=2$. The thin solid line is for 
$\delta\psi_0^{(1,2)}$, and the thick dash-dotted line is $\delta
\psi_1^{(1)}$ and the thick dashed line is $\delta \psi_1^{(2)}$. }
\label{fig:dpsi}
\end{figure}           

The mixing between the transverse and the longitudinal modes can be  quantified
by the angle $\psi$  between the phase vector $\vec{\phi}$
and the wave vector $\vec{\sf k}$ of the perturbation eigenmode,
\begin{equation}
\label{eq:psi}
\psi \equiv \arctan (\frac{\hat{e}_z\cdot\hat{\nabla}\times\vec{\phi}}
{\hat{\nabla}\cdot\vec{\phi}}). 
\end{equation}   
Figure \ref{fig:dpsi} (a) illustrates the relative
positions of $\vec{\phi}$ and  $\vec{\sf k}$ with respect to $\vec{n}_1$.            

Denoting the phase angle for the pure modes by $\psi_0^{(1,2)}$, the pure
transverse mode has $\psi_0^{(1)}=\pi/2$ while $\psi_0^{(2)}=0$
for the pure longitudinal mode. 
From eqs.(\ref{eq:phase:1},\ref{eq:phase:2}) we find that
one eigenvector, $\vec{\phi}_1$, is almost transverse ($\psi\sim \pi/2$) 
whereas the other, $\vec{\phi}_2$, is almost longitudinal ($\psi\sim 0$).  
For each mode we expand the
deviation from $\psi_0^{(1,2)}$ in $\beta$ at the codimension-two point:
\begin{equation}
\label{eq:dpsi}
\delta\psi^{(1,2)}\equiv\psi^{(1,2)}-\psi_0^{(1,2)}=
\delta\psi_0^{(1,2)}(\nu,\theta)+\beta\,\delta\psi_1^{(1,2)}(\nu,\theta) 
+ {\cal O}(\beta^2),      
\end{equation}
where $\theta$ is the angle between ${\sf k}$ and $\hat{n}_1$. 
It turns  out
that in the absence of mean flow the two eigenvectors remain  orthogonal to
each other to fifth-order and therefore  $\delta \psi_0^{(1)}=\delta
\psi_0^{(2)}$.
The deviation $\delta \psi_0^{(1,2)}$ 
depends very little on  $\nu$. The contributions $\delta \psi_1^{(1,2)}$  from the
mean flow decrease with increasing  $\nu$ without changing sign. 
Fig.\ref{fig:dpsi} shows as an example the contributions 
$\delta\psi_{0,1}^{(1,2)}$ to $\delta \psi$ for $\nu=2$. The thin solid lines are
$\delta\psi_0^{(1,2)}$ for the two eigenvectors.  The thick dashed-dotted line is
$\delta\psi_1^{(1)}$  for $\vec{\phi}_1$ and the thick dashed line  is for
$\vec{\phi}_2$. Note that, because $\delta\psi_1^{(1,2)}$ and
$\delta\psi_0^{(1,2)}$ have the same signs and because $\beta$ is negative,
the mean flow always suppresses the mixing between the two  modes. Thus,
the phase perturbations are closer to the pure  transverse/longitudinal modes
for small Prandtl number. This is also found for parameters away from the
codimension-two points (see fig.\ref{fig:phase_angle} below).

\section{General Linear Stability Analysis}
\label{sec:linear02}

\begin{figure}
\begin{center}
\epsfxsize=8.5cm\epsfbox{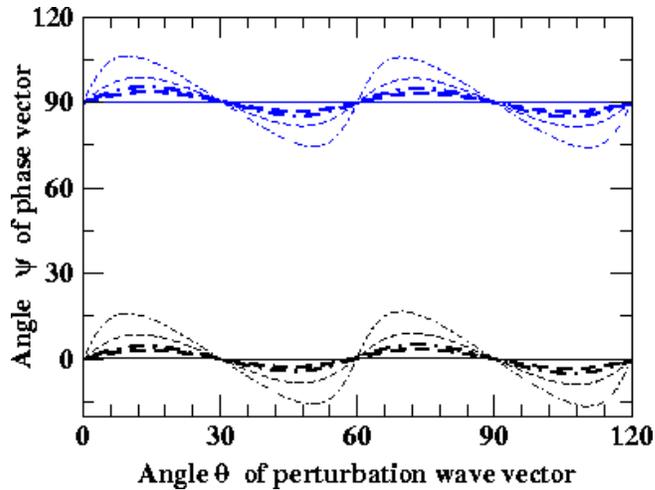}
\end{center}
\caption{Angle $\psi$ (between the phase vector
$\vec{\phi}$ and the perturbation wave vector $\vec{\sf k}$) as a function of
the angle $\theta$ (between $\vec{\sf k}$ and $\vec{n}_1$) for $\mu=0.5$ and
$q=0.5$.   The thin lines are for coupling strength $\beta=0$, and
$\beta=-0.2$ for the thick lines. The solid lines are for ${\sf k}=0$, the
dashed lines for ${\sf k}=0.192$, and the dashed-dotted lines for ${\sf
k}=0.224$. }
\label{fig:phase_angle}
\end{figure}

In this section we present results from solving the linear-stability equations
(\ref{eq:ll1} - \ref{eq:ll6})  numerically for arbitrary perturbation wavenumbers.
We first review and discuss the results for infinite Prandtl number
\cite{LaMe93,SuTs94,EcPe98}. The long-wave analysis of
sec.\ref{sec:linear01} revealed two phase instabilities  associated with
the longitudinal and the transverse phase mode, respectively.  For
finite perturbation wavenumbers these two modes become mixed  and
the eigenmodes of the linear stability problem are not  strictly
longitudinal or transverse any more. Instead they can  be
characterized by the angle $\theta$ between the perturbation
wavevector of the  fastest growing modes and  $\vec{n}_1$. For the
`wide-splitting'  modes   $\theta=n\pi/3$, whereas for the
`narrow-splitting' modes $\theta=\pi/6+n\pi/3$ with $n$  an integer
\cite{SuTs94}. The terminology refers to the fact that the angle
between the relevant most unstable wide-splitting modes is $2\pi/3$,
whereas for the  narrow-splitting modes it is only $\pi/3$
\cite{SuTs94}. 
As the 
perturbation wavenumber goes to 0 the narrow-splitting mode  turns into the
longitudinal mode and the wide-splitting mode  into the transverse phase
mode. This is illustrated in  fig.\ref{fig:phase_angle}, which shows the angle
$\psi$ (eq.(\ref{eq:psi})) 
as a function of the angle $\theta$ between $\vec{\sf k}$ and $\vec{n}_1$.
The thin lines  refer to the case $\beta=0$ discussed here.  
The deviation of the destabilizing modes from these
orientations increases  with $|\vec{\sf k}|$ (cf. fig.\ref{fig:dpsi}). 

\begin{figure}
\centerline{
\epsfxsize=8.5cm\epsfbox{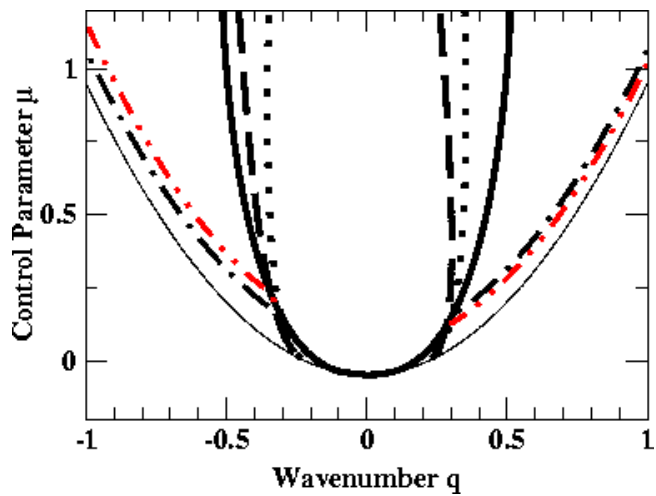}}
\caption{
Cross-over from  narrow-splitting to wide-splitting mode. Along the
dashed-dotted (dashed-double-dotted) line the narrow- and wide-splitting
modes have equal growth rates for $\beta=0$ ($\beta=-0.2$).           
}
\label{fig:sb_summary2}
\end{figure}  

\begin{figure}
\centerline{\epsfxsize=6.5cm\epsfysize=5.5cm\epsfbox{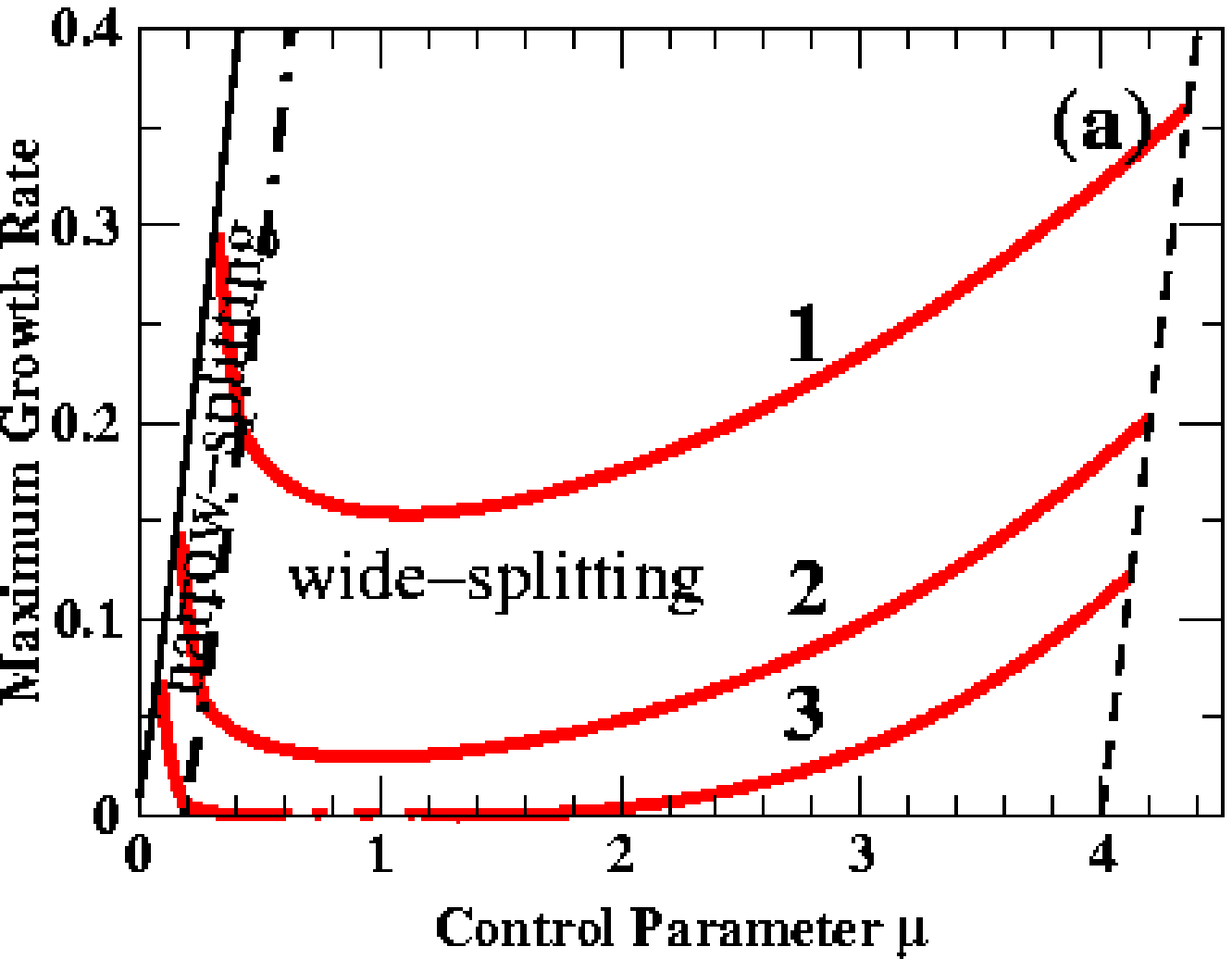}
\hspace{1cm}\epsfxsize=6.5cm\epsfysize=5.5cm\epsfbox{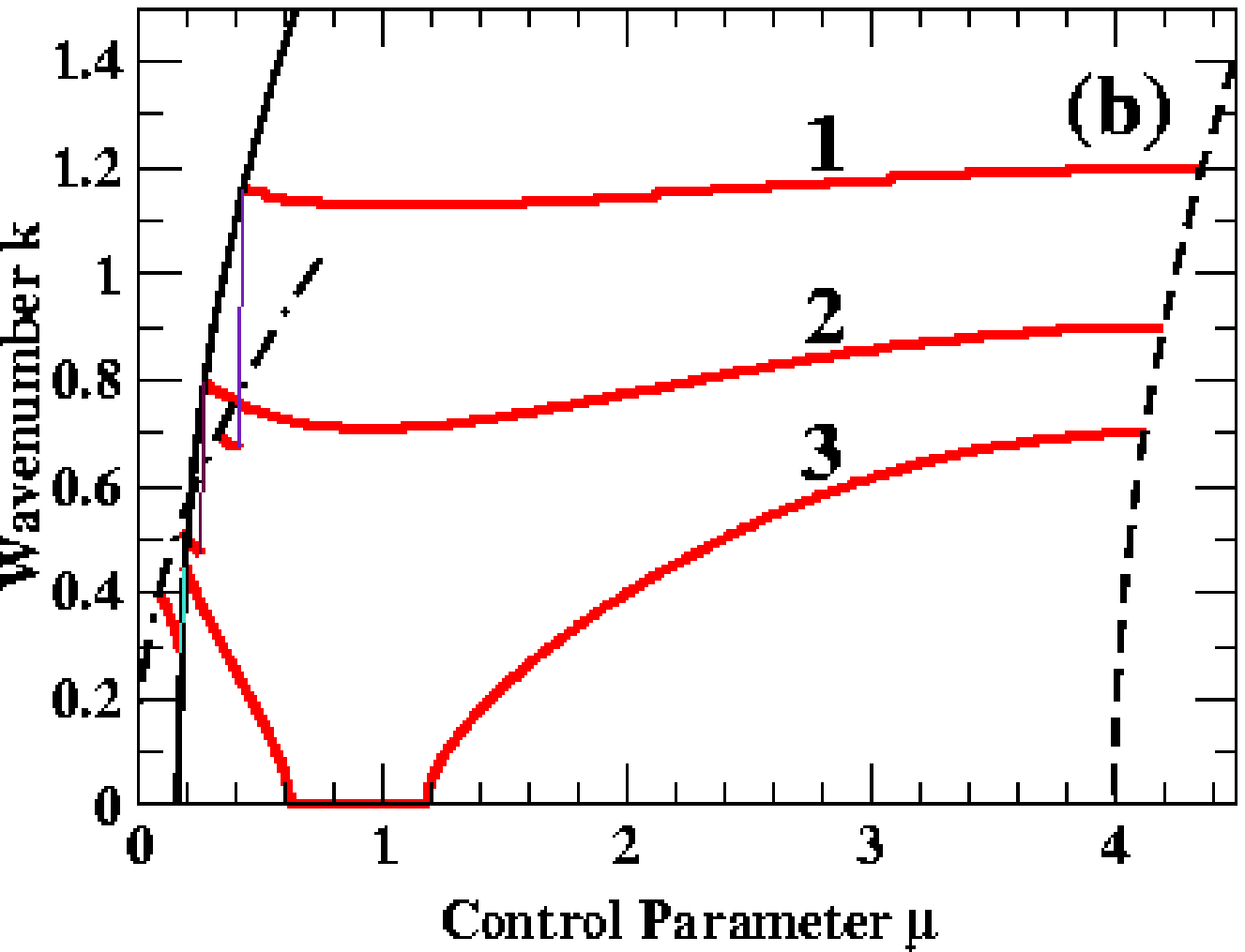}}
\caption{
Results of full linear stability analysis with $\beta=0$.
Panel (a): maximal growth rate as a function of $\mu$ for 
$q=0.6$, $0.45$ and $0.35$ (curves $1$ to $3$).
Panel (b): wavenumber of the fastest growing mode for the corresponding $q$
(curves $1$ to $3$).
}
\label{fig:maxgr_maxk_nomean}
\end{figure}    

Numerical  simulations show that the nonlinear evolution of the narrow- and
of  the wide-splitting modes  leads to  qualitatively different transient
patterns with the  wide-splitting mode inducing noticeably more grain
boundaries  between hexagon patterns of different orientation and 
consequently a larger number of defects than the narrow-splitting mode
\cite{SuTs94}. However, as fig.\ref{fig:no_mf_sb} already suggests, the range
of parameters over  which the longitudinal/narrow-splitting mode  dominates
over the transverse/wide-splitting mode is very small and it is very  difficult
to separate the two modes in numerical simulations \cite{SuTs94}. This is
illustrated further in fig.\ref{fig:sb_summary2} and in
fig.\ref{fig:maxgr_maxk_nomean}. Fig.\ref{fig:sb_summary2} gives  the
cross-over line from the narrow-splitting to the wide-splitting  mode, i.e. the
line along which the growth rates of both modes are  equal.
Fig.\ref{fig:maxgr_maxk_nomean}a shows the dependence of the maximal
growth rate of the dominating mode as a  function of the control parameter
$\mu$ for three values of the  wavenumber.
Fig.\ref{fig:maxgr_maxk_nomean}b shows the dependence of the
wavenumber of the perturbation mode with maximal growth rate. As expected
for a long-wave  instability, the magnitude of the wavenumber of the fastest
growing mode  goes to 0 as the stability limits are reached at  $\mu \approx
0.6$ and  $\mu \approx 1.2$ for $q=0.35$. Although the  relevant
perturbation wavenumber jumps by almost a factor of 2 across the  transition
from the narrow-splitting to the wide-splitting mode (cf. the jump at  $\mu
\approx 0.4$ for $q=0.6$), it  turns out that in simulations this difference is
very hard to identify  because the growth rates of the two modes are so
similar and the  narrow-splitting mode dominates only over such a small
range of $\mu$. 

The mean flow introduces a number of modifications in the stability 
behavior. The linear operator $\cal L$ becomes non-hermitian and
allows complex eigenvalues. It turns out, however, that  in our
calculations complex eigenvalues appear only near the stability
boundary of the mixed mode ($\mu=\mu_{MM}$) and {\it beyond} the
long-wave stability limit given by the  longitudinal mode (cf.
fig.\ref{fig:sb_meanfl_lw}). In numerical  simulations, the transient
oscillations were always accompanied by the  formation of defects
and an eventual transition to rolls.
Possibly,
the mean flow induces more relevant oscillatory  instabilities  in
combination with nonlinear gradient terms,  which also constitute
non-potential terms \cite{Br89,NuNe00}.

\begin{figure}
\centerline{\epsfxsize=6.5cm\epsfysize=6.75cm\epsfbox{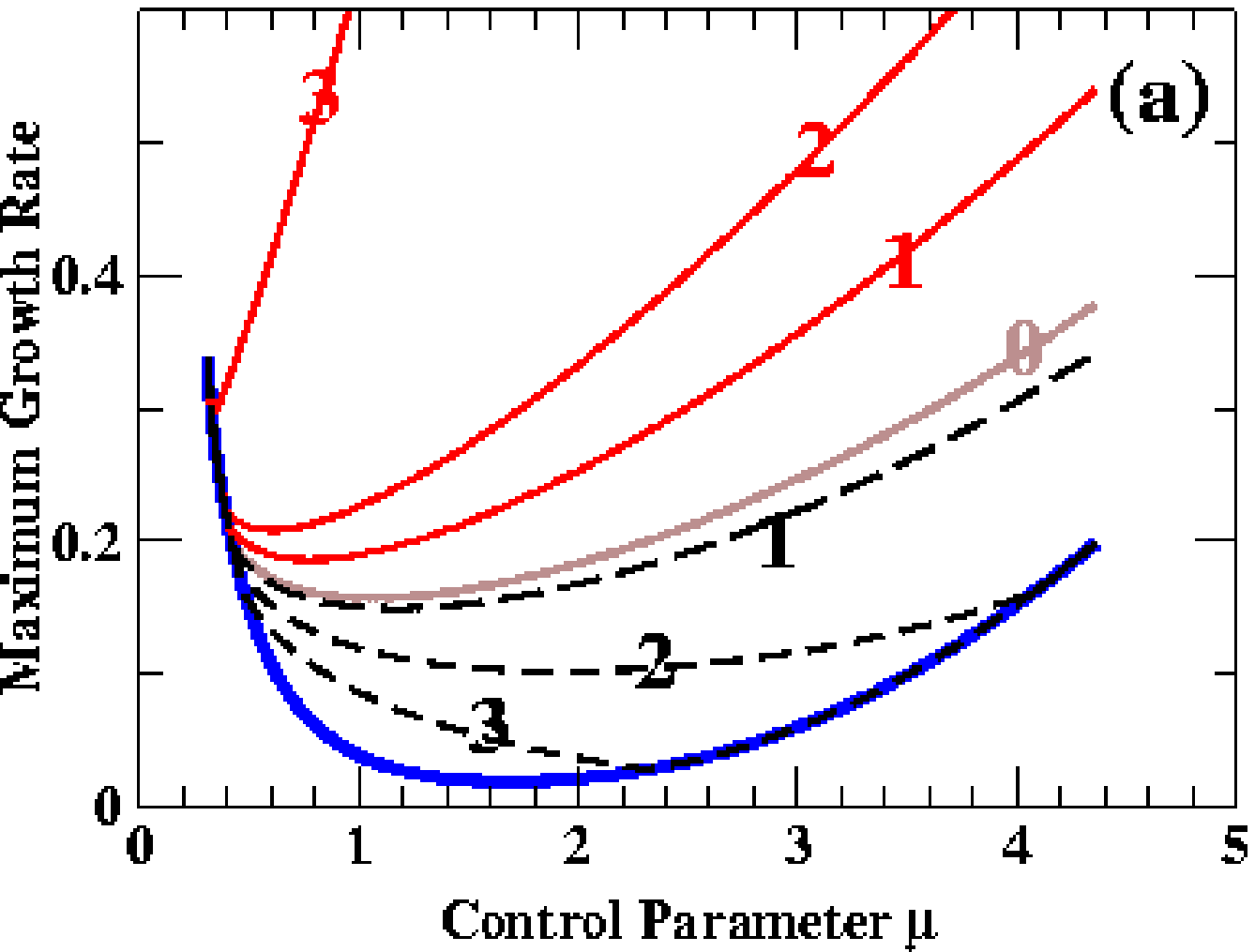}
\hspace{1cm}\epsfxsize=6.5cm\epsfysize=6.0cm\epsfbox{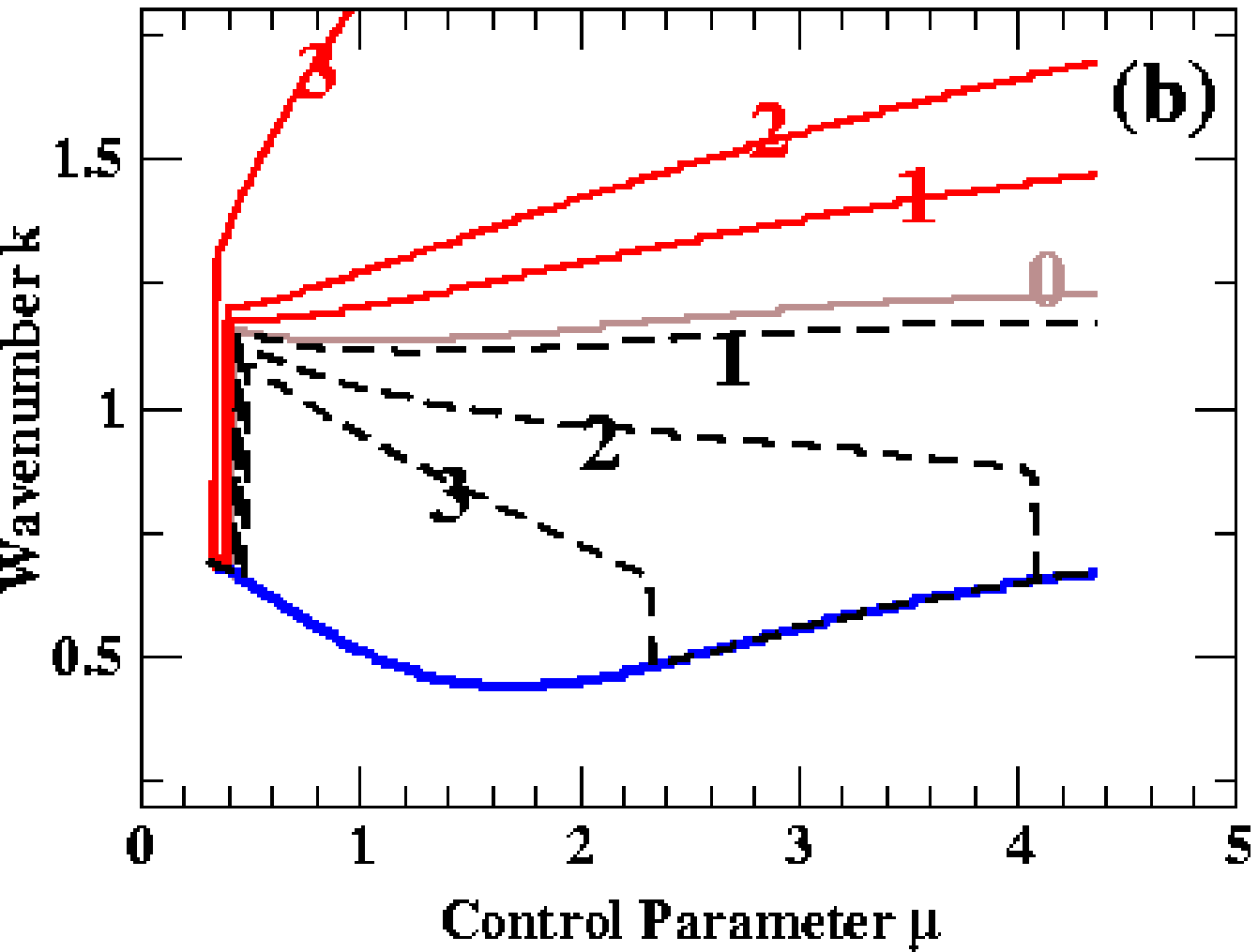}}
\caption{
Results of full linear stability analysis with $q=\pm0.6$.
Solid lines are for $q=0.6$ and dashed lines are for $q=-0.6$.
Panel (a): maximal growth rate as a function of $\mu$.
Panel (b): wavenumber for the fastest growing mode.
From curves $0$ to $3$ $\beta=0$, $-0.1$, $-0.2$, $-0.3$, respectively.
}
\label{fig:maxgr_maxk_mean}
\end{figure}

The mixing between the longitudinal and the transverse mode for finite 
values of ${\sf k}$ persists in  the presence of mean flow. It is, however,
strongly suppressed, consistent with our analytical results for the vicinity 
of the codimension-two point (cf. eq.(\ref{eq:dpsi})).
This is illustrated in fig.\ref{fig:phase_angle}, 
where the thick lines give  the angle
$\psi$ of the phase vector for $\beta=-0.2$. The  difference  between
the growth rates of the narrow- and of the wide-splitting modes is still  quite
small for small values of $\mu$, as it is in the case without  mean flow. This is
illustrated in fig.\ref{fig:maxgr_maxk_mean}, which  shows the maximal growth
rates and the associated wavenumbers for  $q=+0.6$ and for $q=-0.6$ 
using solid and dashed lines, respectively, for different coupling strengths.
For larger values of $\mu$, however, even moderate values of the coupling
enhance the  difference noticeably. Thus, for $\beta q<0$ the  wide-splitting
mode becomes much more  dominant. For $\beta q>0$ the narrow-splitting
mode, which in the absence of mean flow is only relevant in a very narrow
range of  the control parameter near the saddle-node bifurcation,  becomes
dominant again above a critical value of $\mu$. This reflects the asymmetry of the
stability boundaries induced by the mean flow as shown in
fig.\ref{fig:sb_meanfl_lw}a,b. The critical  value of $\mu$  decreases with 
decreasing Prandtl number, and below Prandtl number $\sim 2.6$ $(\beta=-0.4)$, 
the transverse mode is preempted by the longitudinal mode over the 
whole range of $\mu$. This suggests again that for sufficiently small
Prandtl numbers it should be experimentally possible to study  the two
instability modes separately.     

\section{Numerical simulation}
\label{sec:num}
We now present results from numerical simulations of
eqs.(\ref{eq:mf1},\ref{eq:mf2}) using a parallel pseudo-spectral code. In
sec.\ref{sec:num01} we compare the nonlinear evolution of the transverse
and longitudinal side-band instabilities. We then discuss the impact of the
mean flow on the competition between rolls and hexagons. In
sec.\ref{sec:num02} we study the motion of penta-hepta defects in the
presence of mean flow. These results shed some light on the persistence of grain 
boundaries in disordered hexagons patterns.


\subsection{Nonlinear Evolution of the Side-band Instabilities}
\label{sec:num01}              

{\bf Transverse vs. longitudinal modes.} 

\noindent Without mean flow the nonlinear
evolution of the two side-band instabilities of hexagon patterns has been
studied numerically in some detail by Sushchik and Tsimring \cite{SuTs94}. 
They find that both instabilities lead to the formation of penta-hepta  defects
and subsequently to disordered hexagon patterns. When only   the
wide-splitting perturbations are growing the emerging patterns have many grain
boundaries  separating domains of hexagons with different orientation.  These
disordered patterns persisted for a long time. When the  narrow-splitting
instability is dominant only few such domains appear  and a regular hexagon
pattern with a different wavenumber is restored  within a relatively short time.
Sushchik and Tsimring note that numerical simulations in  the regime in which
the narrow-splitting mode is the only destabilizing  mode are very difficult
and only present results for parameter values  for which also  the
wide-splitting mode is destabilizing albeit to a lesser degree.  These  runs
correspond to a quench deep into the unstable regime beyond the 
dashed-dotted line in fig.\ref{fig:no_mf_sb} ($q=0.56$ and  $\mu=0.6$). The
authors associate the larger number of grain boundaries in the first case with
the large angle between the components of the wide-splitting mode. A 
detailed comparison is difficult since no simulations are available in  which
only the  narrow-splitting mode is active. 

\begin{figure}
\centerline{\epsfxsize=4.0cm\epsfbox{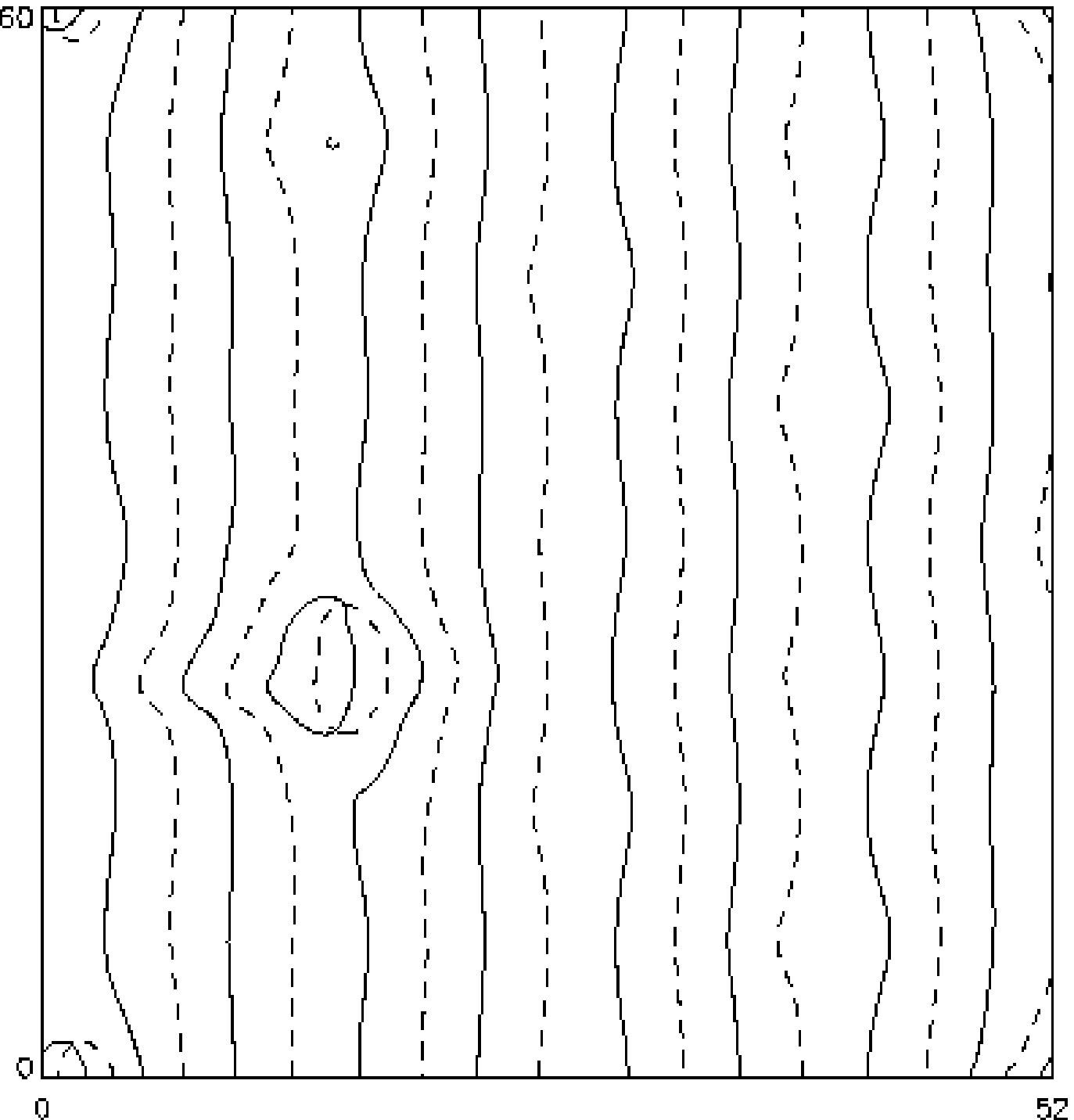}
            \epsfxsize=4.0cm\epsfbox{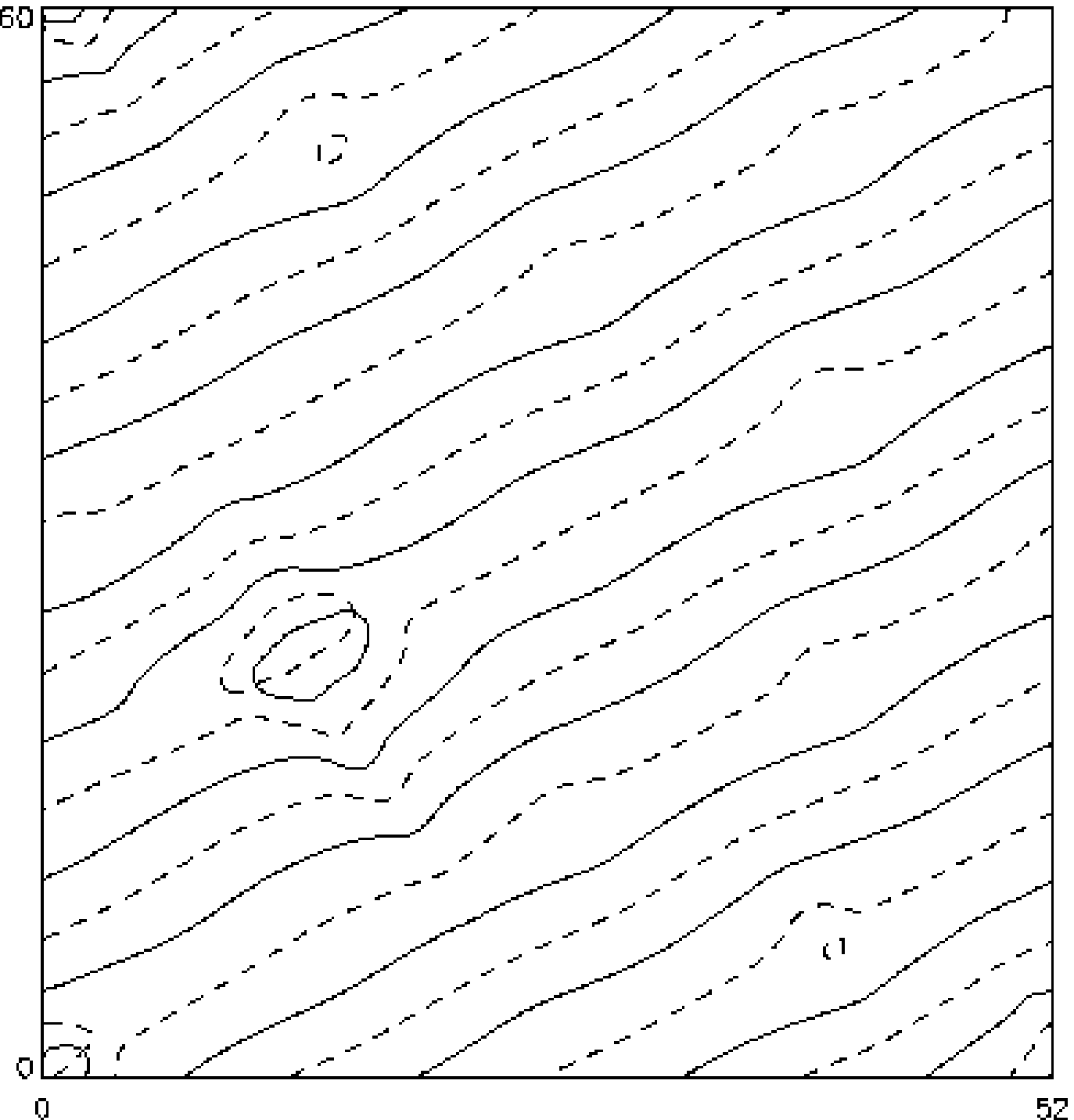}
	    \epsfxsize=4.0cm\epsfbox{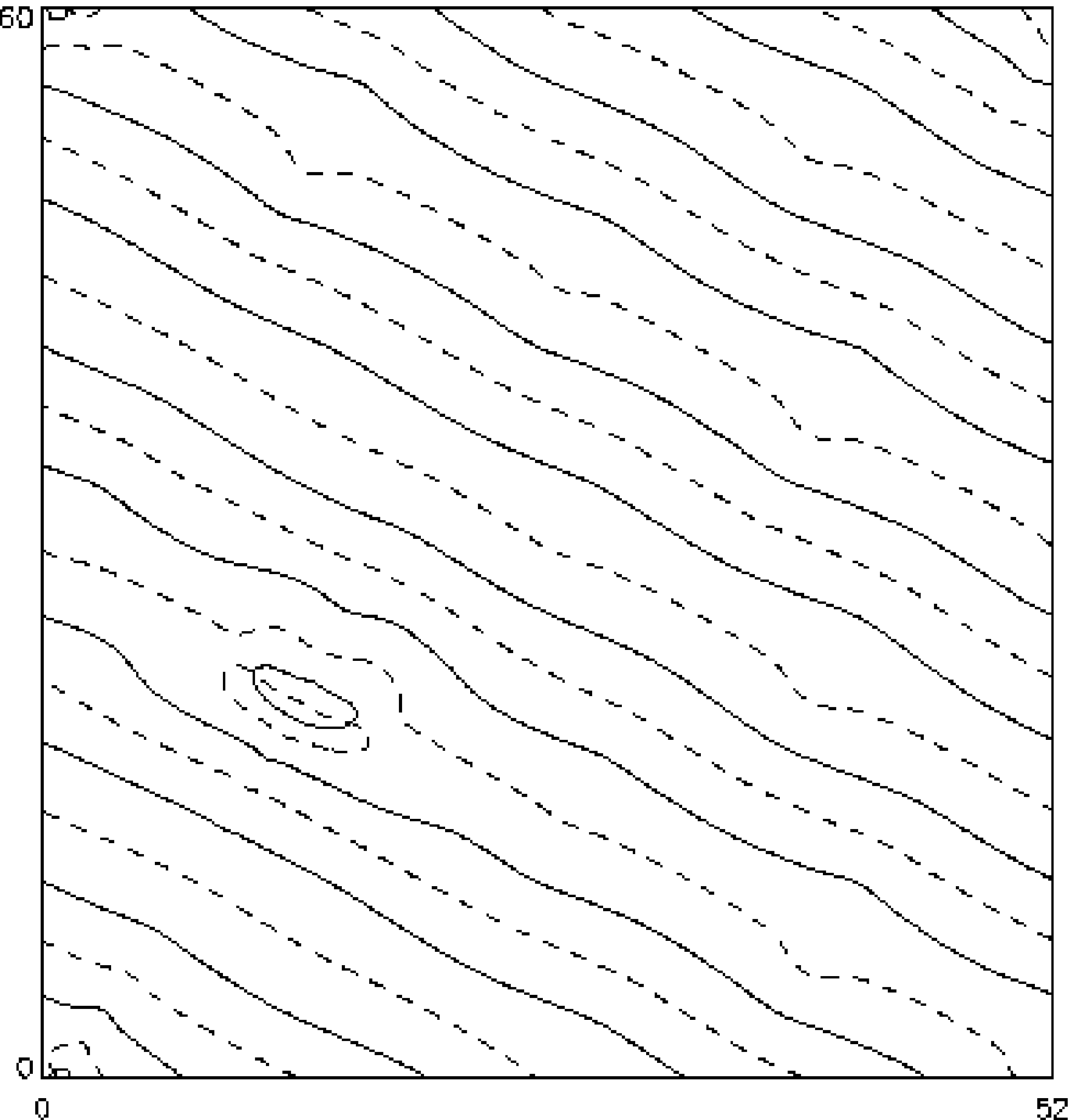}}
\centerline{
            \epsfxsize=4.0cm\epsfbox{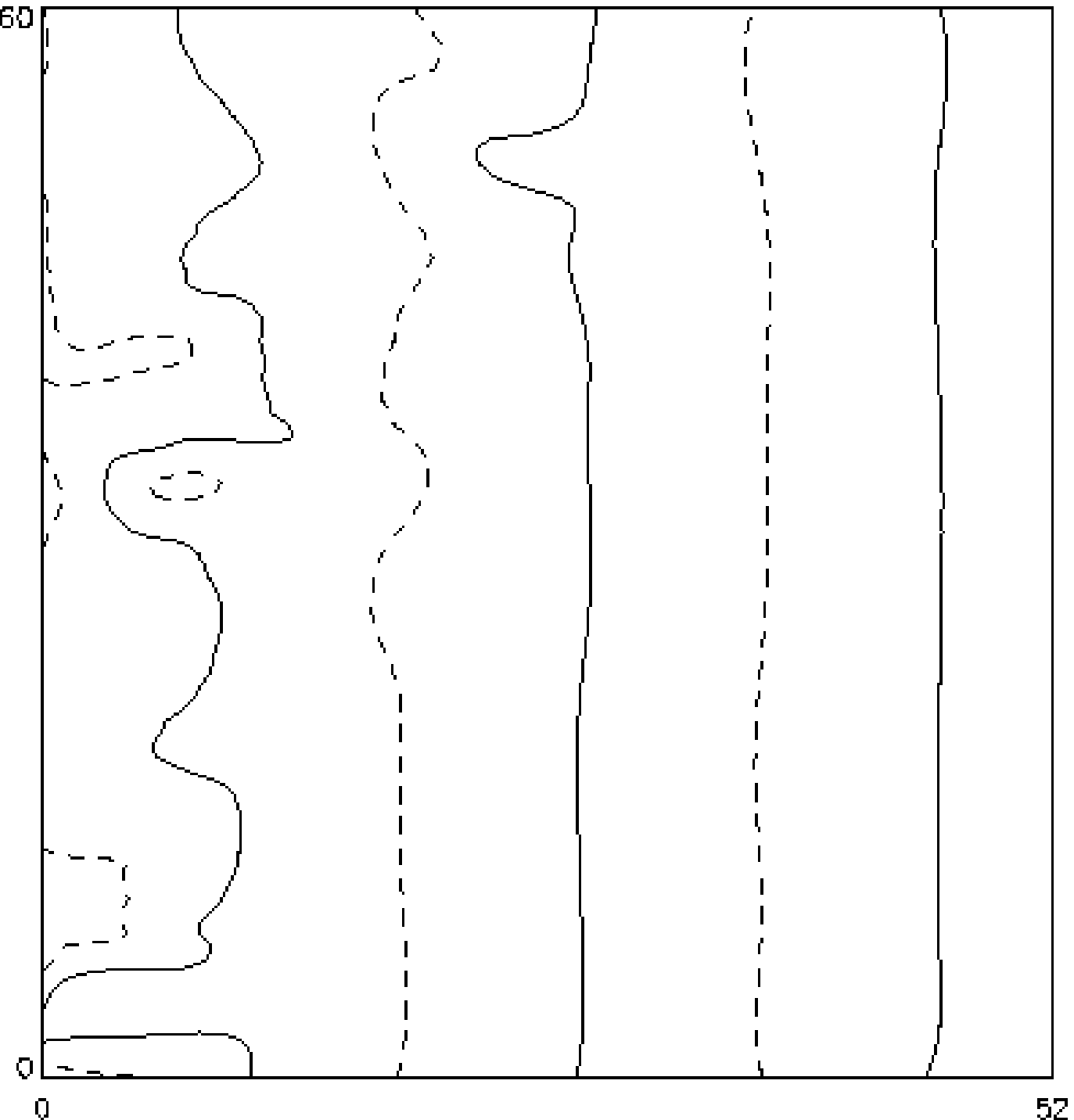}
            \epsfxsize=4.0cm\epsfbox{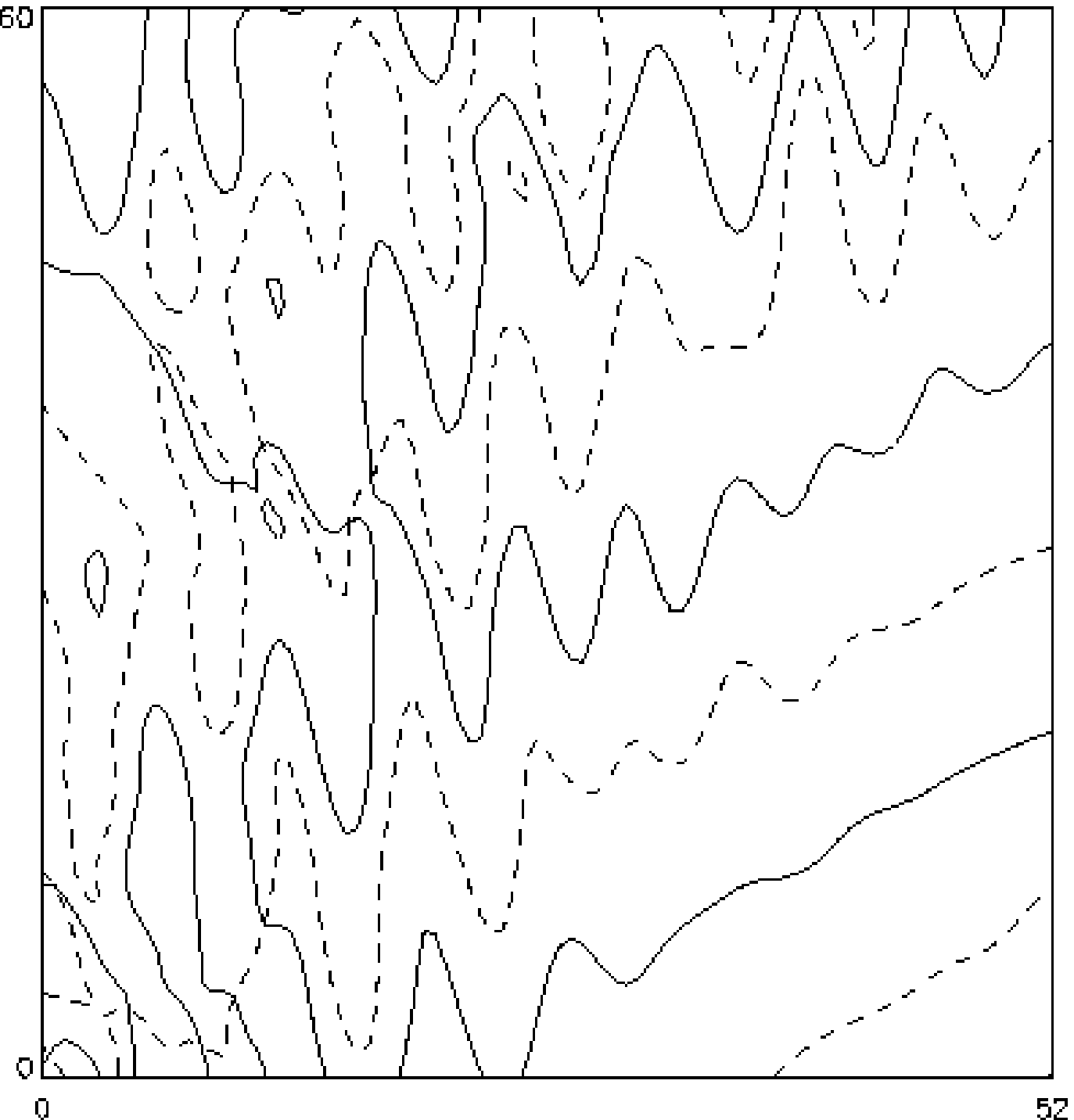}
            \epsfxsize=4.0cm\epsfbox{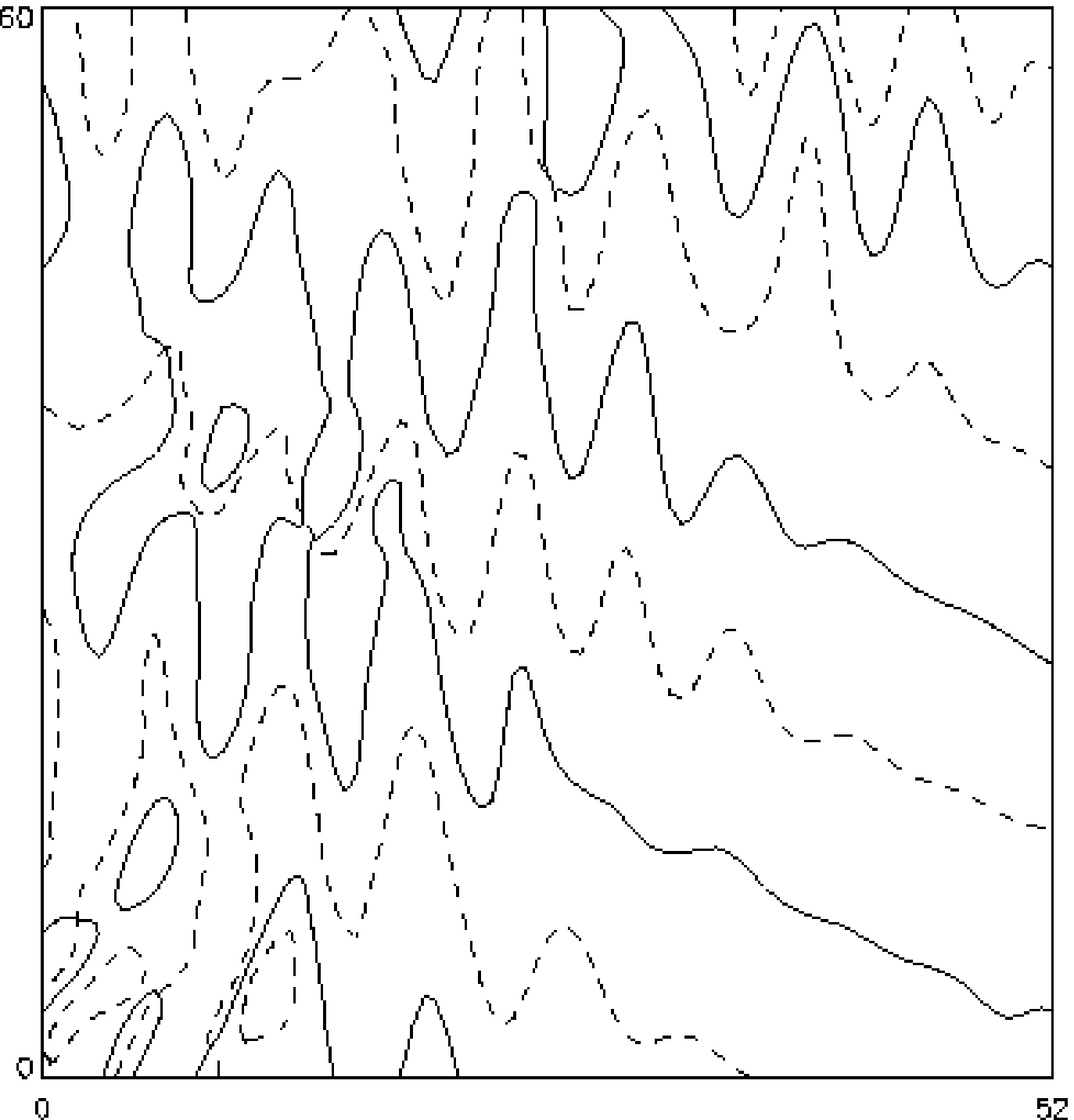}}
\caption{Zero contour lines of the amplitudes 
         ($A_1$ to $A_3$ from left to right)
         for $\beta=-3$ and $\mu=0.5$.  Solid lines are zero contours of the
         real parts, and the dashed lines are the imaginary parts.
         The top three panels are the longitudinal modes 
         ($q=-0.48$) and the bottom
         three panels are the transverse modes ($q=0.17$) 
         at $t=100$ in the simulations.  Only a quarter of the domain is shown.
}
\label{fig:curl_div_evolution}
\end{figure}

\begin{figure}
\centerline{
\epsfxsize=8.5cm\epsfbox{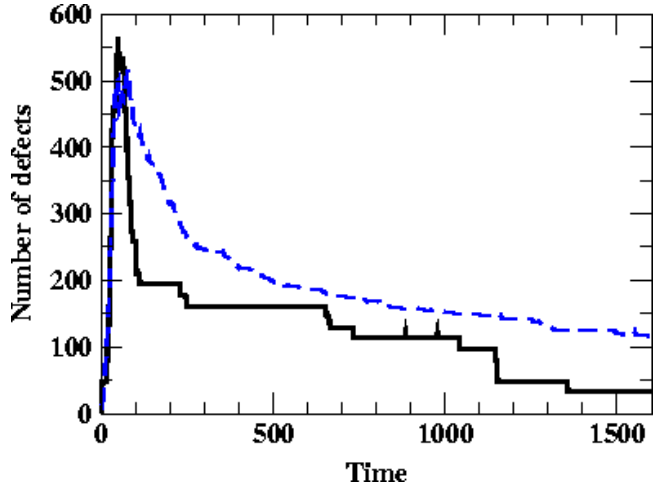}}
\caption{Number of defects as a function of time for the longitudinal
case (solid line) and transverse case (dashed line).
}
\label{fig:curl_div_def}
\end{figure}

\begin{figure}
\centerline{
            \epsfxsize=6.0cm\epsfbox{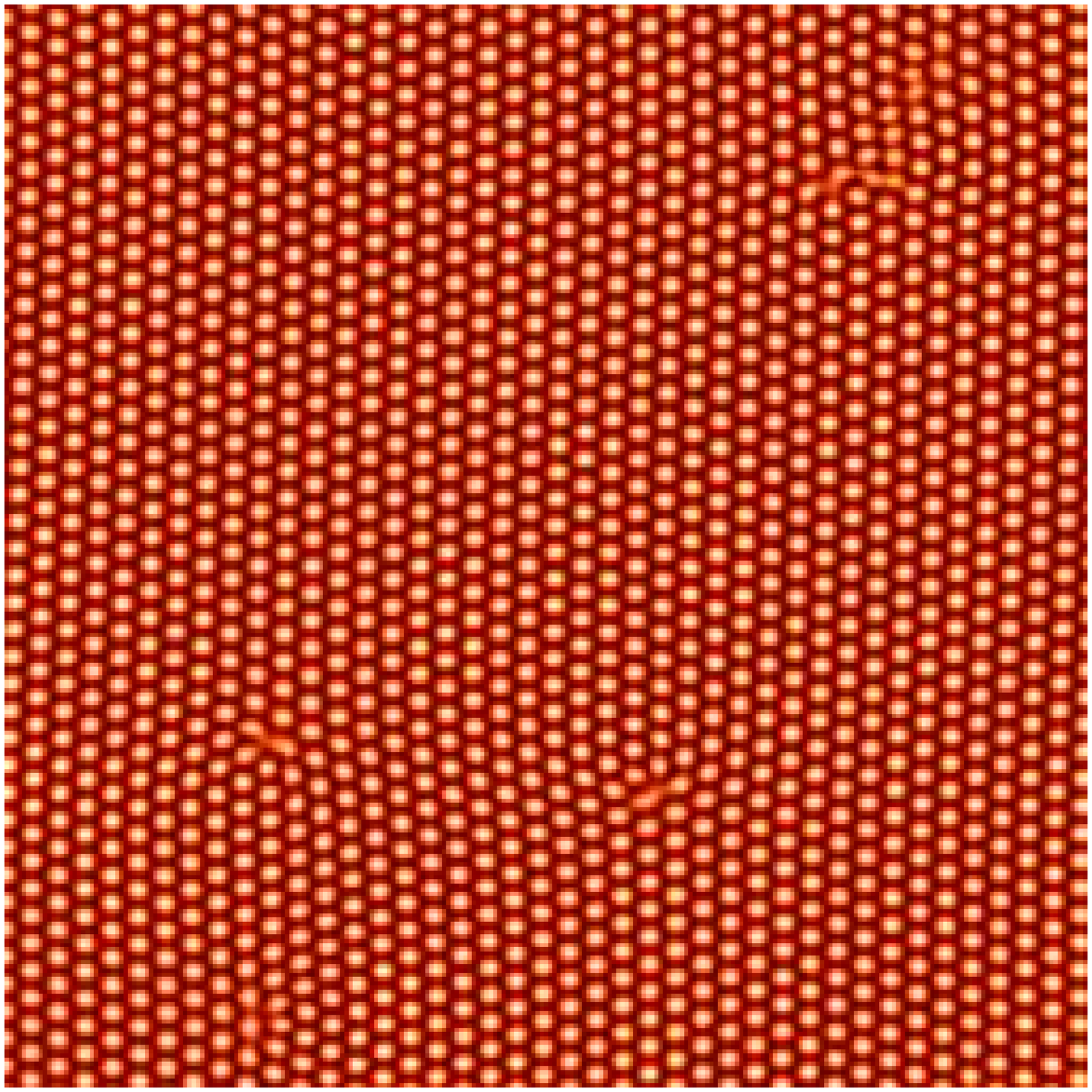}
\hspace{1cm}\epsfxsize=6.0cm\epsfbox{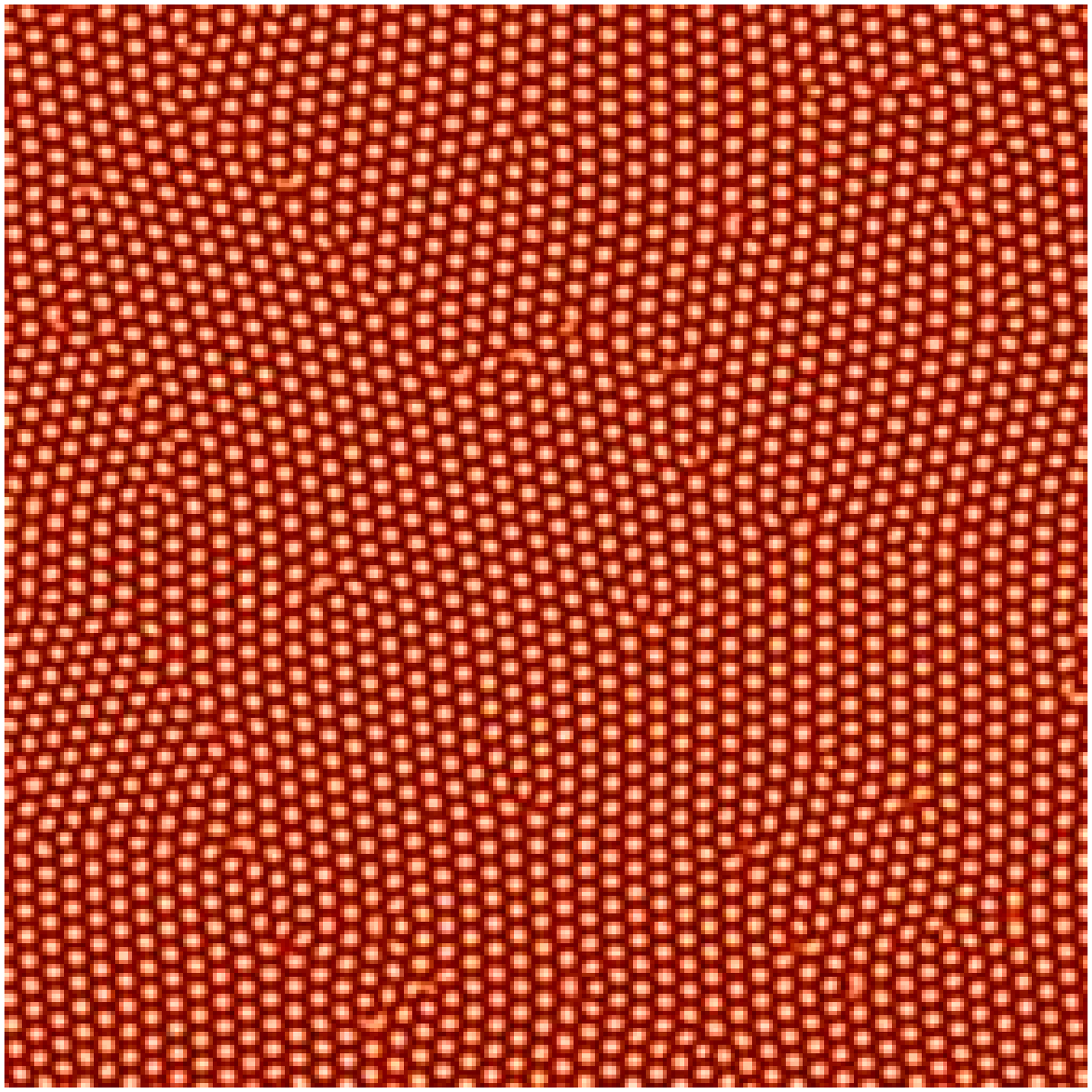}}
\caption{Reconstructed temperature 
for the longitudinal mode (left panel, $q=-0.48$)
and the transverse mode (right panel, $q=0.17$) at $t=1750$.
}
\label{fig:curl_div_evolution1}
\end{figure}

As discussed in sec.\ref{sec:linear01} and sec.\ref{sec:linear02}, the
mean flow couples only to the transverse/wide-splitting  instability
and for  sufficiently small Prandtl number suppresses it completely for
$\beta q  > 0$. This makes it possible to compare the two instabilities
in detail  under comparable conditions.  
Figs.\ref{fig:curl_div_evolution}-\ref{fig:curl_div_evolution1} show a
comparison of the evolution of the  two instabilities for $\mu=0.5$,
$\nu=2$, and $\beta=-3$. The system size is $L=104$  and the
numerical resolution is $256\times 256$.
The wavenumbers of the initial, slightly perturbed regular hexagon
patterns  are chosen carefully to obtain the same linear growth rates
for both  modes. In particular, for $q=-0.48$ ($q=0.17$) only the
longitudinal  (transverse) mode is destabilizing (cf.
fig.\ref{fig:sb_meanfl_lw}).

Fig.\ref{fig:curl_div_evolution} shows the early evolution of the 
instabilities in terms of the contour lines of the real and imaginary 
parts of the three amplitudes $A_i$, $i=1,2,3$. While the longitudinal 
mode shown in the top three panels induces compressions and dilations 
of the pattern, the transverse mode shears the amplitudes. Both lead to 
the formation of defects. However, the defects induced by the 
longitudinal mode are typically aligned along the rolls and form where 
the bulges are maximal. 

The temporal evolution of the number of defects 
is shown in fig.\ref{fig:curl_div_def}. For both modes, we define $t=0$ 
when the first defect is formed. Both modes having the same growth rate 
it is not surprising that in both cases the defect number grows on a 
similar time scale. In fact, both reach roughly the same number of 
defects at about the same time $t \approx 50$. In both cases the 
subsequent ordering of the pattern appears to occur in two stages, 
characterized by an initial rapid annihilation of defect pairs and a 
later much slower phase. In the longitudinal case the defect number 
decreases in large steps, which are associated with the annihilation of 
a string of defects roughly aligned with the rolls,
whereas no such steps are visible in the transverse case. 
The overall decay is substantially slower for the pattern induced by 
the transverse instability. This had also been found in the absence of 
mean flow \cite{SuTs94}. 

Snapshots of the reconstructed patterns at the final time  $t=1750$ are shown
in fig.\ref{fig:curl_div_evolution1}. Only a quarter of the whole system is  shown.
While in the longitudinal the defect density  case is very low and the 
defects are essentially isolated from each other, in the transverse 
case most of the defects are part of grain boundaries. 
         
\begin{figure}
\centerline{
\epsfxsize=6.0cm\epsfbox{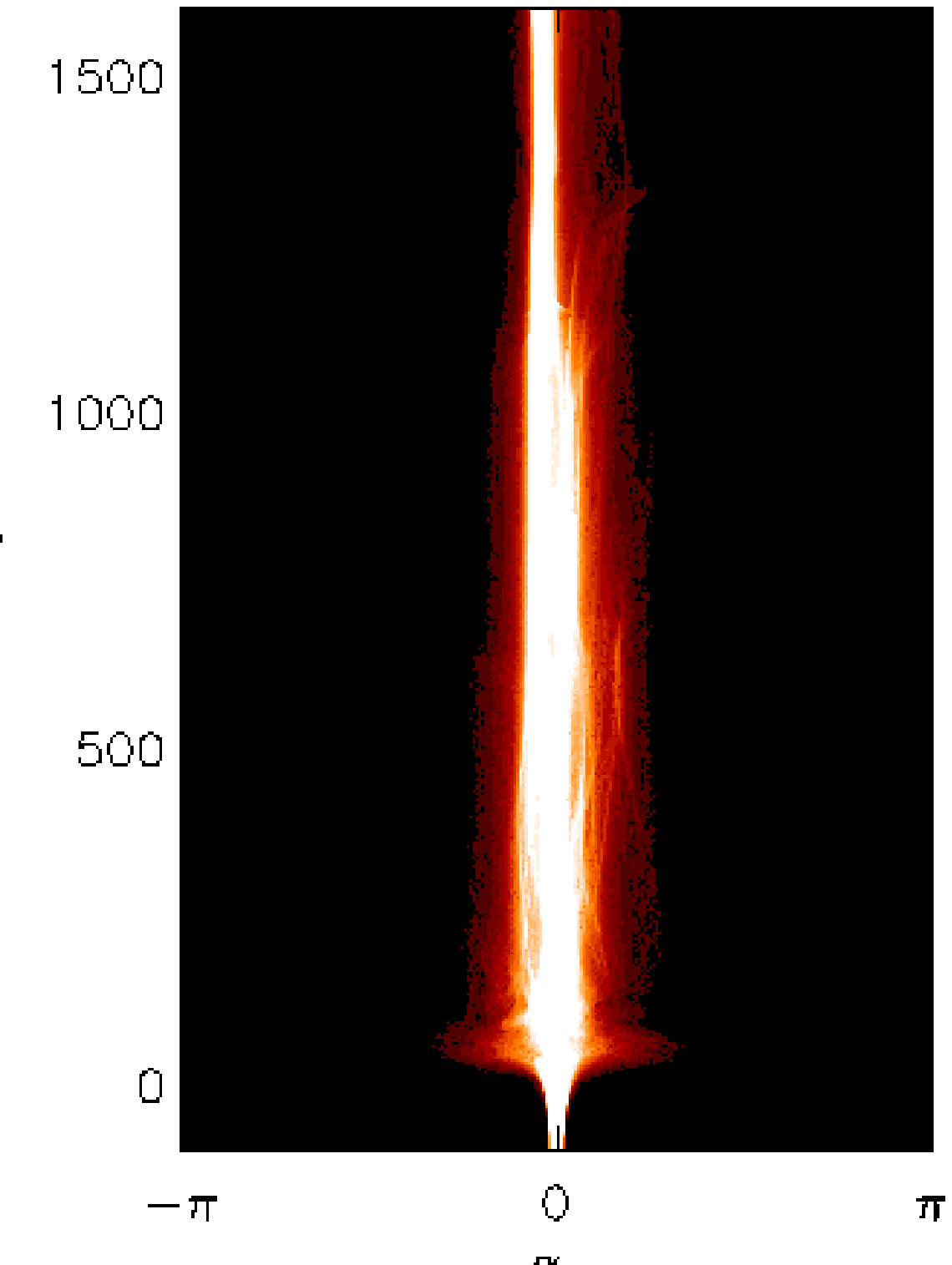}\\
\epsfxsize=6.0cm\epsfbox{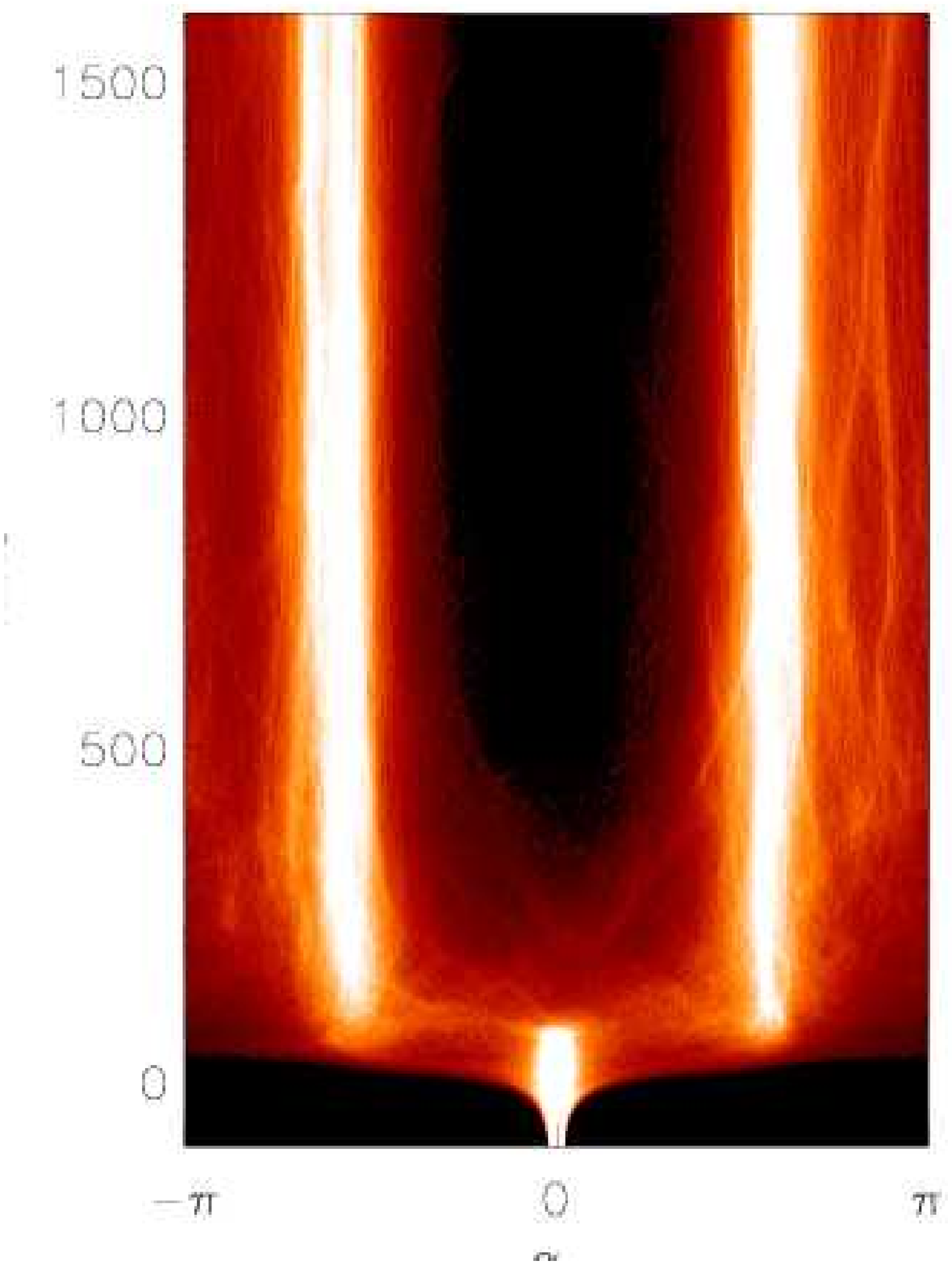}
}
\caption{
Evolution of the probability distribution 
of $\alpha_1$: the angle between $\vec{k}_1$
and $\hat{n}_1=\hat{x}$ 
for both longitudinal (left) and transverse (right).
}
\label{fig:curl_div_ori}
\end{figure}

To quantify the evolution of the amount of disorder in the orientation of the 
hexagons we determine the orientation of the local wavevector $q_j$ relative
to the roll direction $\hat{n}_j$ as defined by the angle $\alpha_j$:
\begin{equation}
\label{eq:local_wn}
\alpha_j \equiv \arctan
\left(\frac{\vec{q}_j\cdot\hat{\tau}_j}{\vec{q}_j\cdot\hat{n}_j}\right),\;\;
\vec{q}_j\equiv \Re \left(\frac{-i\vec{\nabla}A_j}{A_j} \right)\;\; {\mbox{ for $j=1$, $2$, $3$}}.
\end{equation}
In figure \ref{fig:curl_div_ori} we plot the probability distribution function
(PDF) of the orientation angle  $\alpha_1$ as a function of time for the cases
shown in figures \ref{fig:curl_div_evolution} and \ref{fig:curl_div_def}.  Again,
time $t=0$ is where the first dislocation occurs, and we have truncated  the
large peaks (around $\alpha_1=0$) at early times so that more structures can
be discerned at later times.   Similar PDFs are found for the other two
amplitudes as well.  For the longitudinal mode the PDF is centered around
$\alpha_1=0$ from the beginning to the end, with the peak broadening
around $t=0$ when the first few defects appear. Thus, for all times  there is
only a single domain and the hexagons remain essentially aligned with their
initial orientation.  In the transverse case, however, the initial peak at
$\alpha_1=0$ quickly decays and gives way to two peaks of comparable
size. This occurs around the time when the maximum number of defects is
reached. The bi-modal PDF indicates that the  transverse mode
predominantly induces hexagons of two different orientations, which then
co-exist for a long time. 
    
{\bf Hexagons vs. Rolls.} 

\noindent 
In the absence of mean flow, the competition
between uniform hexagons and rolls is governed by the energy difference
between them. For a given $q$,  hexagons are more stable than  rolls for
$\mu$ lower than a threshold value $\mu_{th}(q)$,  at which the rolls and
hexagons have the same free  energy \cite{SuTs94},  while rolls are
energetically favored  above $\mu_{th}$. This boundary is indicated by the
dashed line in figure \ref{fig:no_mf_sb}. Below the dashed line, outside the
stability balloon, the unstable transverse and longitudinal modes grow and
evolve  towards a hexagon of different wavenumber.  Above this line, rolls
appear during the transients and eventually replace the hexagon.

\begin{figure}
\centerline{\epsfxsize=6.5cm\epsfbox{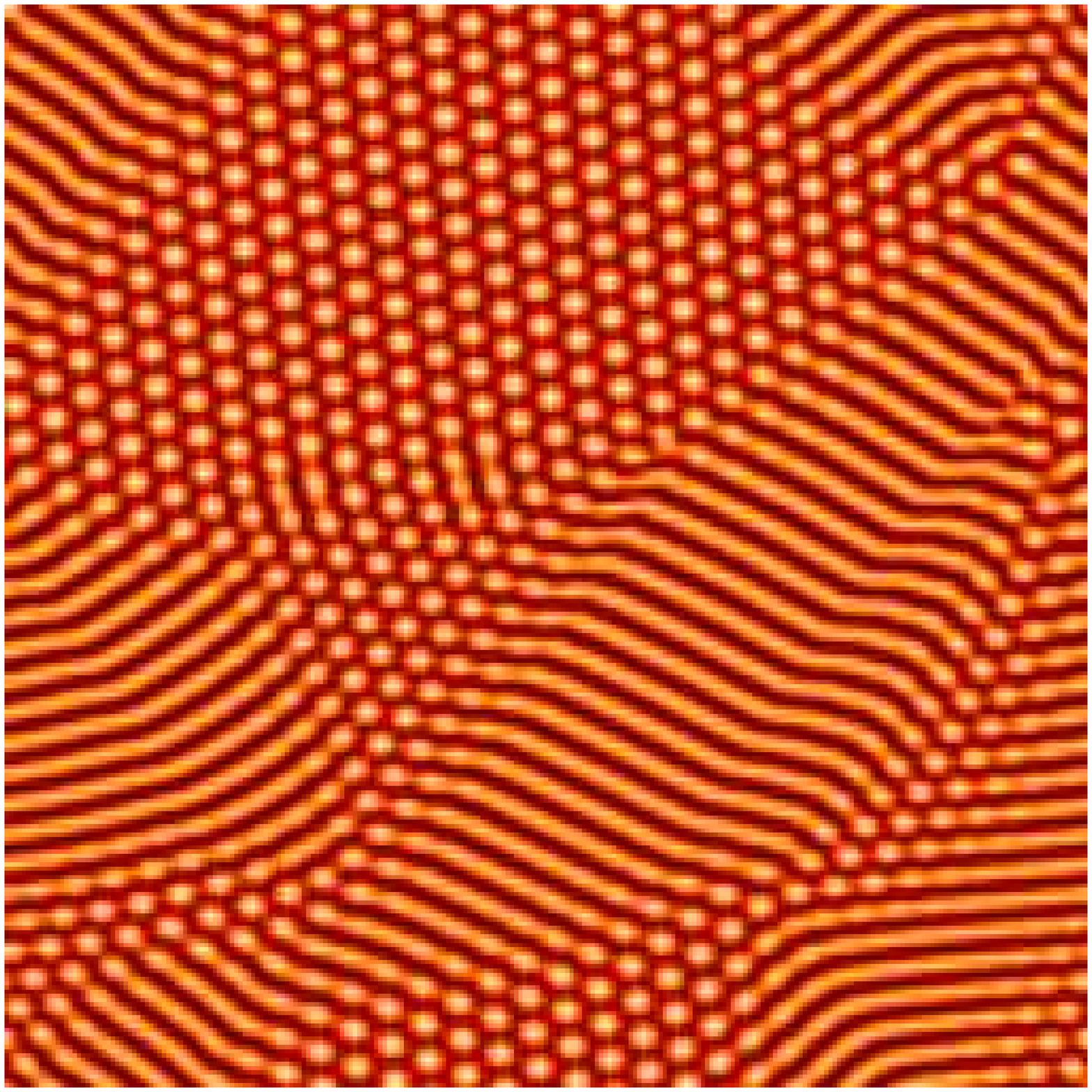}
\hspace{1cm}\epsfxsize=6.5cm\epsfbox{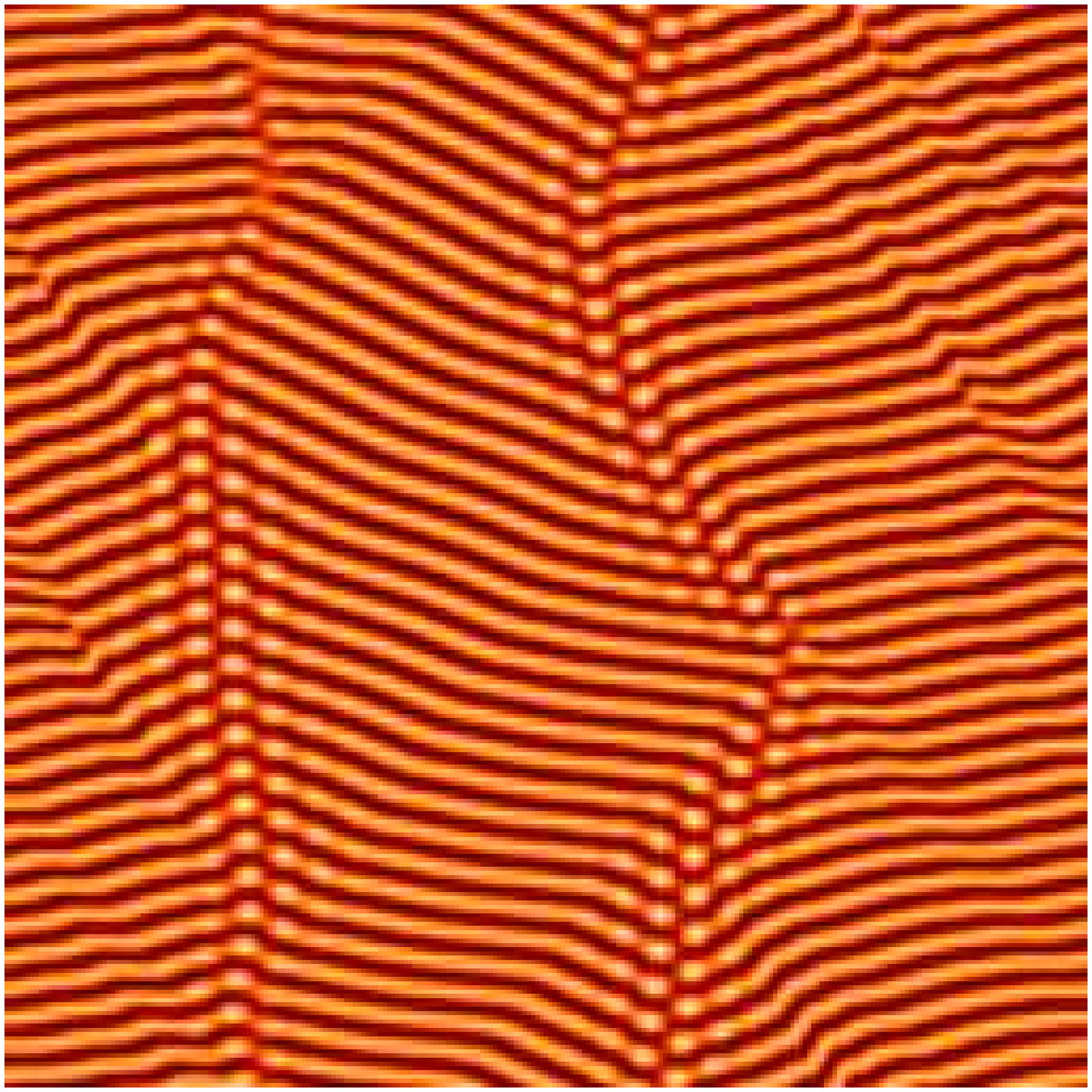}}
\caption{
Reconstructed patterns from simulating equations (\ref{eq:mf1},\ref{eq:mf2})
for $\beta=-0.1$ and $q=0.6$. 
On the left, $\mu=1.425$ at $t=2000$,
and $\mu=1.45$ on the right panel at the same time.
}
\label{fig:roll_hex_01}
\end{figure}

Due to the lack of a Lyapunov functional  for eqs.(\ref{eq:mf1},\ref{eq:mf2}),
we  resort to numerical simulations to locate the boundary between hexagon
and roll in the presence of mean flow.  Table \ref{table1} 
lists $\mu_{th}$ for $q=\pm 0.6$ for different values of $\beta$. 
For example, for $\beta=-0.1$ (Prandtl number $=10$)
and $q=0.6$ we find rolls as the final state 
  at $\mu=1.45$ (figure
\ref{fig:roll_hex_01}(a)) while a mixed state of rolls and hexagons is found at
$\mu=1.425$  (figure \ref{fig:roll_hex_01}(b)). Thus, the transition value in
$\mu$ at $q=0.6$ for $\beta=-0.1$ is  between $1.425$ and $1.45$. 
The enhanced instability of the transverse mode for $q=0.6$ leads 
apparently to an earlier transition to rolls when the Prandtl number is
decreased. At this point it is not clear why the converse is not the 
case for $q=-0.6$. 

\begin{table}
\begin{center}
\begin{tabular}{ccrr}
 $\beta$ & $\Pr$         & $q=0.6$     & $q=-0.6$  \\
\hline
 $0$     & $\infty$      & $\mu_{th}=1.5$   & $\mu_{th}=1.5$ \\
 $-0.1$  & $10$          & $1.43$  & $1.5$ \\
 $-0.2$  & $5$           & $1.35$  & $1.5$ \\
 $-1.24$ & $1$           & $1.25$  & $1.5$ \\
 $-3$    & $0.5$         & $1.1$   & $1.5$ \\
\end{tabular}
\end{center}
\caption{
Competition between rolls and hexagons
for $q=\pm 0.6$ for different Prandtl numbers.
}
\label{table1}
\end{table}

\subsection{Effect of mean flow on  motion of defects}
\label{sec:num02}
In this section we study the effect of the mean flow on the motion of 
a penta-hepta defect (PHD), which is a bound state of two dislocations in two
of the three amplitudes. For large Prandtl number, where the mean flow is
negligible, there exists a semi-closed form solution for stationary
penta-hepta defects for  hexagons at the band-center ($q=0$)
\cite{Ts96}.
We study the general case with mean
flow numerically by embedding two PHDs in the system as initial conditions
and measure their  velocity as a function of $\beta$ and $q$ for fixed
$\mu=1$ and $\nu=2$.

The numerical resolution is  $256\times 256$ in a system of size 
$L=400$ and the time step is fixed at $0.1$ for most of the results
presented in this subsection. To satisfy the periodic boundary
conditions,  we place two PHDs of charges  $(0,+1,-1)$ and
$(0,-1,+1)$ in the computational domain, i.e. each PHD consists of two
dislocations of opposite  charges in the amplitudes $A_2$ and $A_3$.
We also apply a circular ramp at $R=0.4L$, beyond which the phase
is set to constant \cite{Ts95}.  The interaction between pairs of PHD is
characterized by the number  $N\equiv \sum^{3}_{i=1}
\delta_j^1\delta_j^2$, where $\delta_j^{1(2)}$ is the charge of the first 
(second) PHD in the $j$th amplitude \cite{Ts96}. In our simulations $N=-2$ and
the PHDs attract each other.  Their interaction   decreases with
distance and we find that it becomes negligible for distances larger
than $300$ (for $\beta=0$ and $q=0$ the interaction-induced
velocity is then below $v=0.001$). Thus in the following, we place
the two PHDs at least at a distance of $300$ apart in the initial
configuration for the velocity measurement of individual PHD. 

\begin{figure}
\centerline{
\epsfxsize=8.5cm\epsfbox{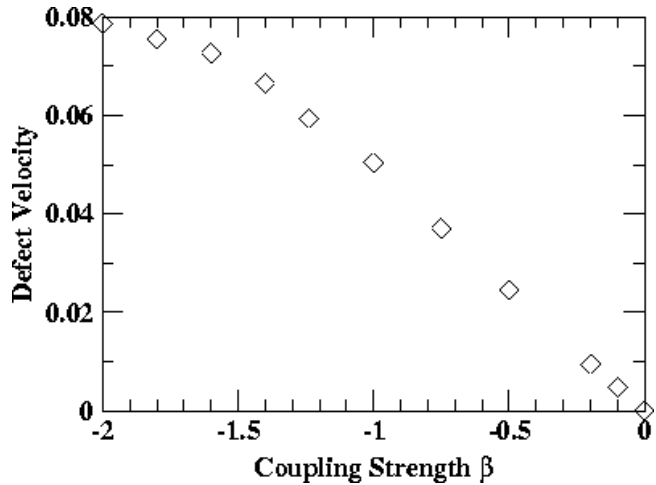}}
\caption[]{
Defect velocity as a function of the coupling strength $\beta$  for $q=0$ and
$\mu=1$. As in the case of rolls \cite{Wh76}, for Prandtl number $\Pr \ge 0.5$
the defect speed scales as $\sim \Pr^{-1}$ (cf.  fig.\ref{fig:beta_pr}). }
\label{fig:def_beta_q0}
\end{figure}

In the absence of mean flow, each independent PHD is found to move at a
constant velocity, which vanishes at $q=0$ \cite{Ts95,Ts96}. In the presence
of mean flow, we also find that isolated defects move at a constant velocity. It
is shown in figure \ref{fig:def_beta_q0} as a function of $\beta$ for $q=0$.    
The range of the linear scaling with respect to the coupling strength indicates
that, for small $\beta$, the contribution of the mean flow to the PHD  velocity
is purely additive via the advective term $i \beta A(\hat{\tau}\cdot\nabla)Q$
and that the amplitudes $A_i$ of the defect solution are only weakly affected
by the mean flow. However, as $|\beta|$ increases, the defect
velocity deviates significantly from the linear scaling. 
This is analogous to the effect of mean flow on dislocations
in roll pattern \cite{BoPepriv}.
\begin{figure}
\centerline{
\epsfxsize=6.5cm\epsfbox{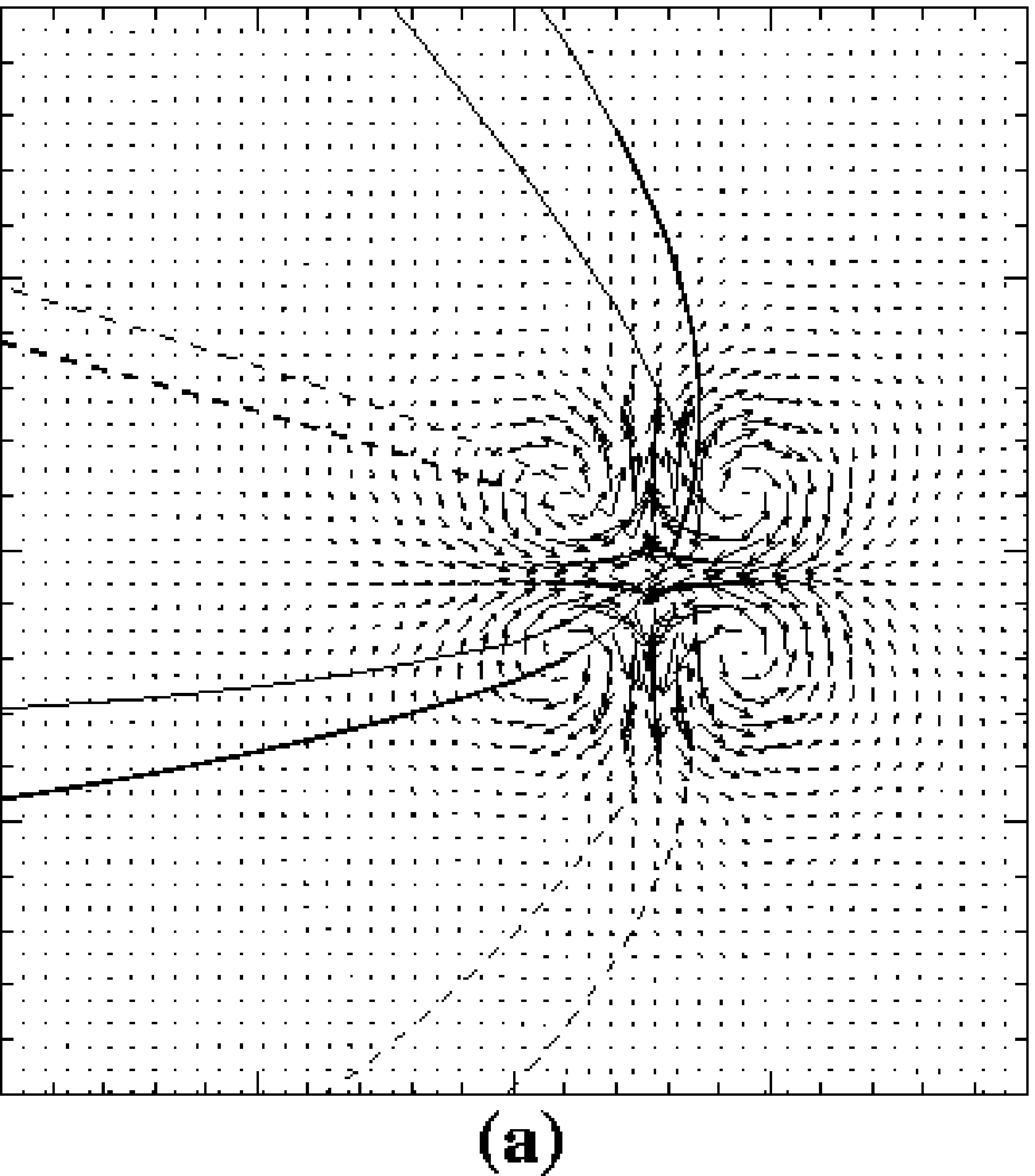}
\hspace{1.0cm}
\epsfxsize=6.5cm\epsfbox{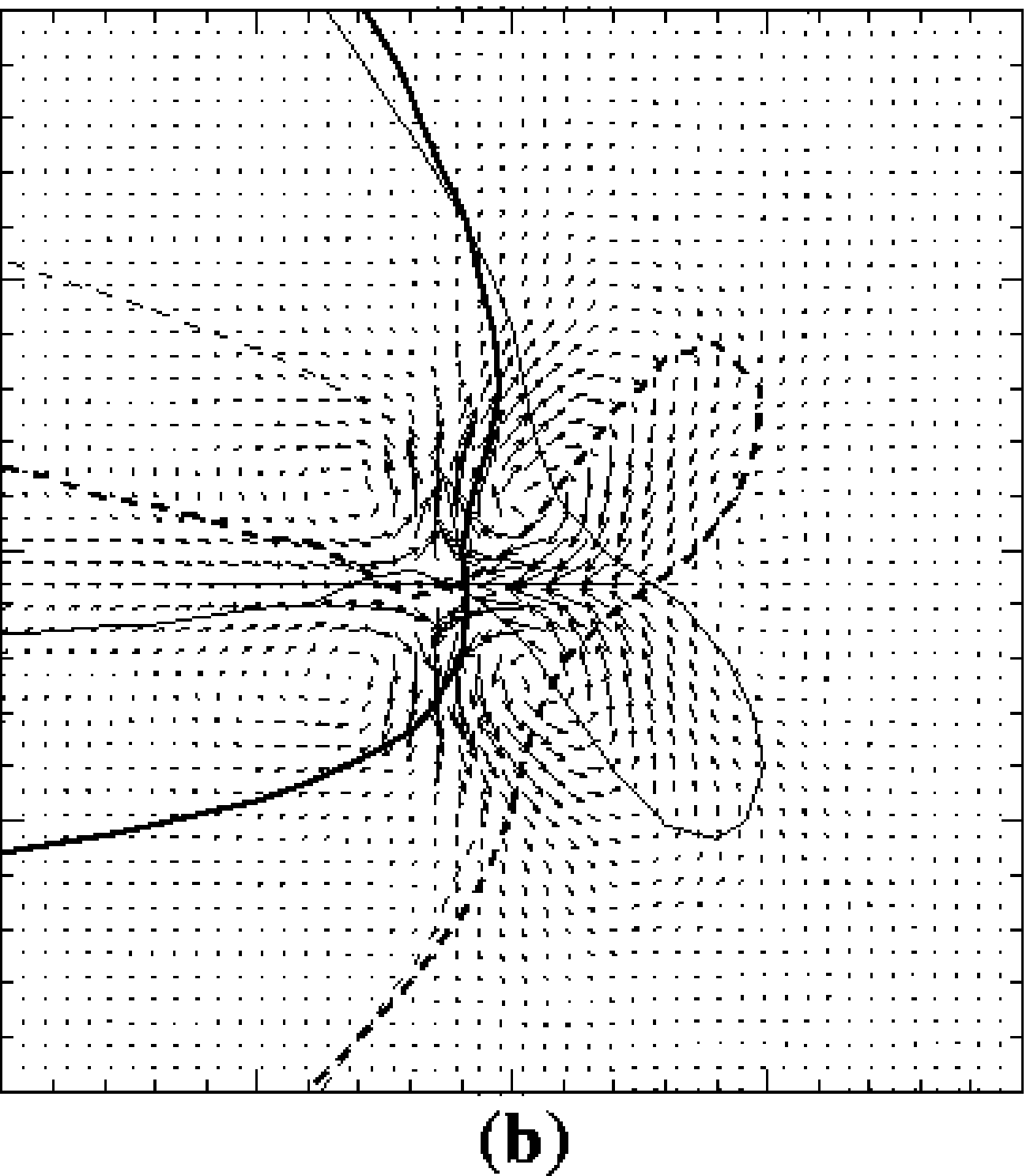}}
\caption{
Mean flow structure around a PHD $(q=0)$: Panel (a) for $\beta=-0.2$
and panel(b) for $\beta=-1.4$.
The solid lines (dash lines) are
the zero contour lines for the real (imaginary)
parts of the second amplitude are shown.
}
\label{fig:def_mf}
\end{figure}    
 Also, for larger $|\beta|$ the mean flow
structure becomes distorted near the defect  as shown in figure
\ref{fig:def_mf}.  The mean flow consists of two pairs of vortices, and is
almost zero away from the defect. While for $\beta=-0.2$ (left panel)
the two vortex pairs are of comparable strength, the pair on the right
is much stronger than that on the left when $\beta$ is decreased to
$\beta=-1.4$ (right panel) This change occurs smoothly in $\beta$.
When the Prandtl number is decreased further so that $\beta$ is
below $-2$ the stability  limit comes very close to the background
wavenumber $q=0$ of the pattern and the PHD's triggers the  formation
of additional defects in their vicinity.  

\begin{figure}
\centerline{\epsfxsize=6.5cm\epsfbox{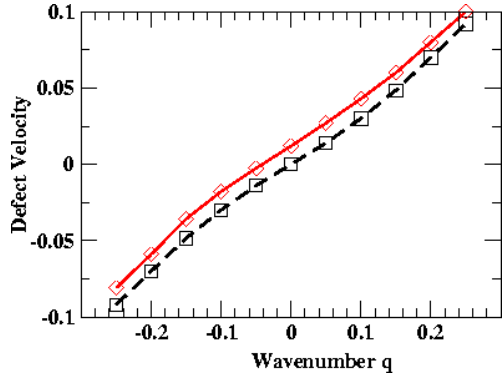}}
\caption{Defect velocity as a function of wavenumber $q$ for $\mu=1$, $\beta=0$ (dashed line)
and $\beta=-0.2$ (solid line).
}
\label{fig:def_q}
\end{figure}

The defect velocity also depends on the wavevectors $\vec{q}_i$  of the
three modes making up the hexagon pattern \cite{Ts95}. More specifically,
within equations (\ref{eq:mf1},\ref{eq:mf2}) it
depends only on  the projections $\vec{q}_i\cdot \hat{n}_i$.  Figure
\ref{fig:def_q} shows the defect velocity as a function of 
$q\equiv\vec{q}_i\cdot \hat{n}_i $ for $\beta=0$ (dashed line) and
$\beta=-0.2$ (solid line). The mean flow shifts  the defect velocity to more
positive values for all $q$, implying that the wavenumber  $q_{st}$ at
which the defect remains motionless is shifted from $q_{st}=0$ to negative
values.  For situations in which the evolution from disordered to more ordered 
patterns is dominated by defects this suggests that the wavenumber of  the
final state is in general not the critical wavenumber or that with  the maximal
growth rate but depends on the  Prandtl number through the mean flow.

\begin{figure}
\centerline{\epsfxsize=5.0cm\epsfysize=5.5cm\epsfbox{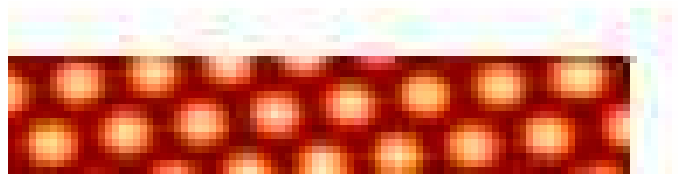}
\hspace{1cm}\epsfxsize=5.0cm\epsfysize=5.5cm\epsfbox{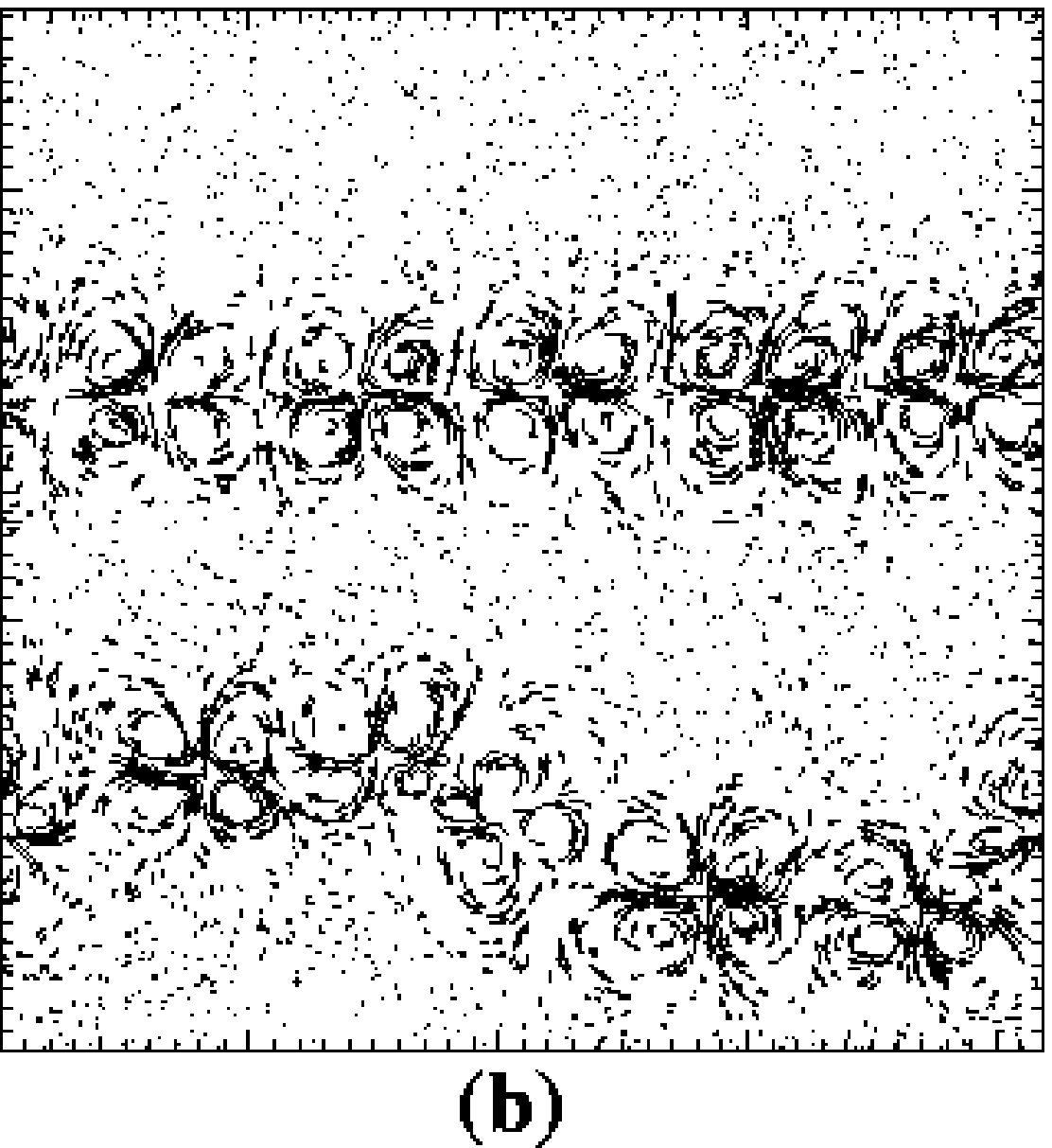}
}
\caption{
Panel(a): Reconstructed hexagon pattern ($q_c=4$) at $t\sim 14000$ for
$\beta=-1.24$ (Prandtl number $\sim 1$) and $\mu=1.2$. The initial 
condition was an ordered, unstable hexagon pattern with  $q=0.60$.
Panel(b): Mean flow of the state shown in (a).
}
\label{grainboundary_1}
\end{figure}

In various simulations of the nonlinear evolution of the instabilities of the 
hexagon pattern we found disordered states characterized by grain 
boundaries between domains of hexagons of different orientation that  moved
exceedingly slowly. An example of such a long-lived disordered  state and
the associated mean flow is shown in  
fig.\ref{grainboundary_1}a,b, where
the spatial structure of such long-lived grain boundaries and the
corresponding mean flow is depicted for $\mu=1.2$, $q=0.6$ and 
$\beta=-1.24$. Here the
resolution is $128\times 128$ for a system size of $L=42$.

\begin{figure}
\centerline{
\epsfxsize=5.0cm\epsfbox{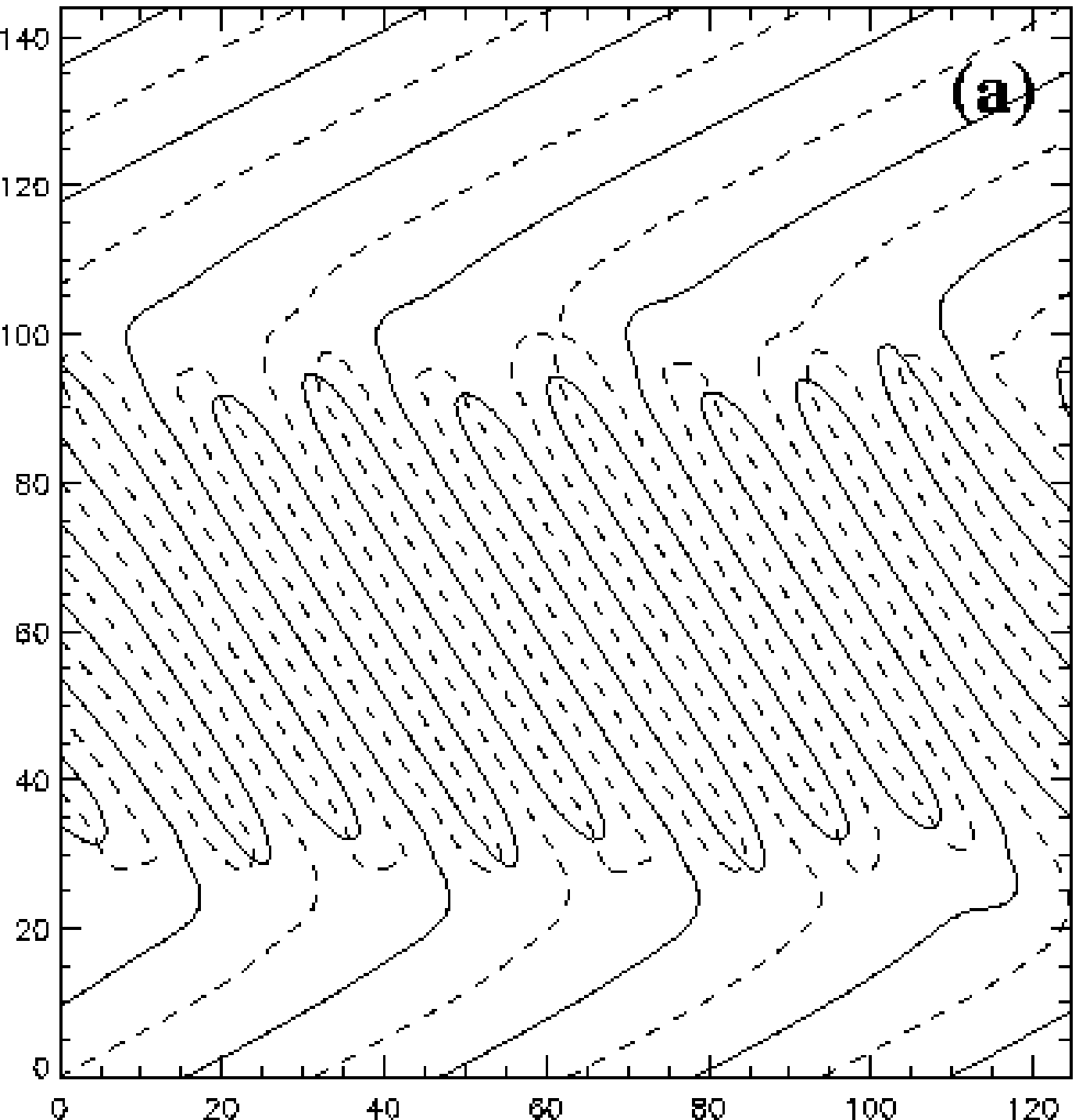}
\hspace{0.2cm}\epsfxsize=5.0cm\epsfbox{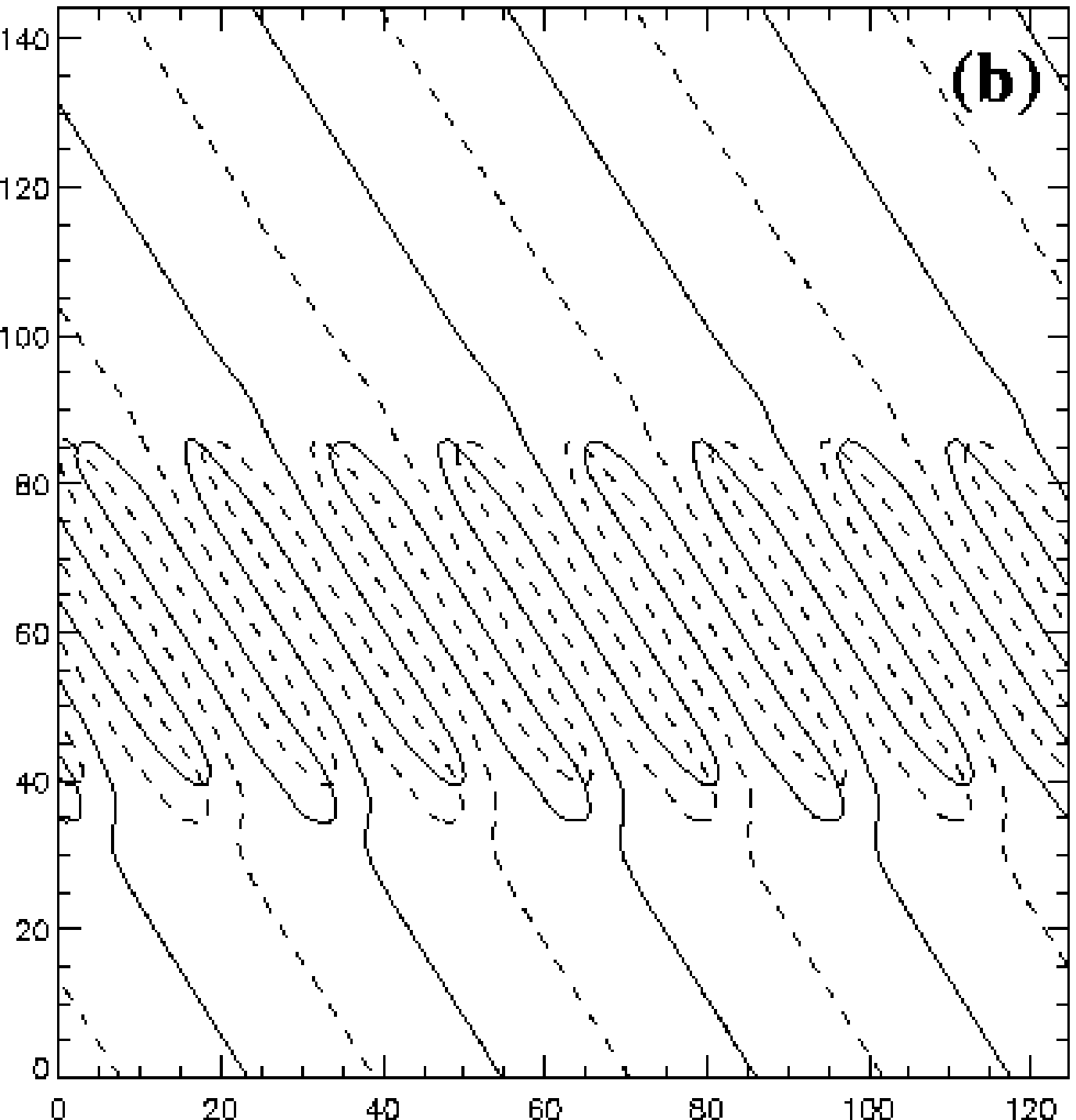}
\hspace{0.2cm}\epsfxsize=5.0cm\epsfbox{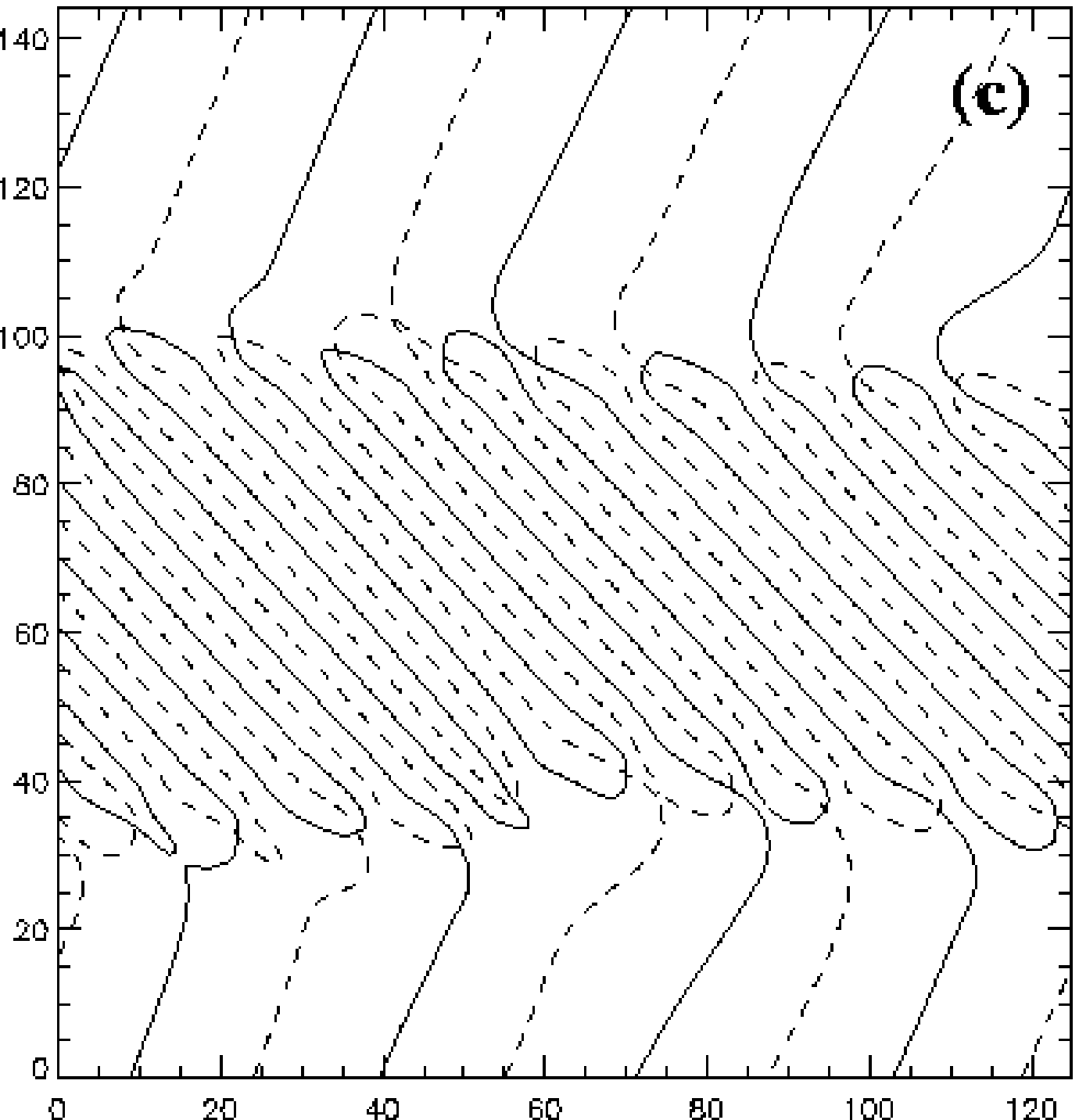}}
\caption{
Zero contour lines of the real and the imaginary part of $A_2$.  Panel (a) is
for the initial conditions, and panels (b) and (c) are snapshots at
$t=80000$  for $\beta=0$ and $\beta=-2.0$, respectively. }
\label{fig:defects_01}
\end{figure}
\begin{figure}
\centerline{
\epsfxsize=5.0cm\epsfbox{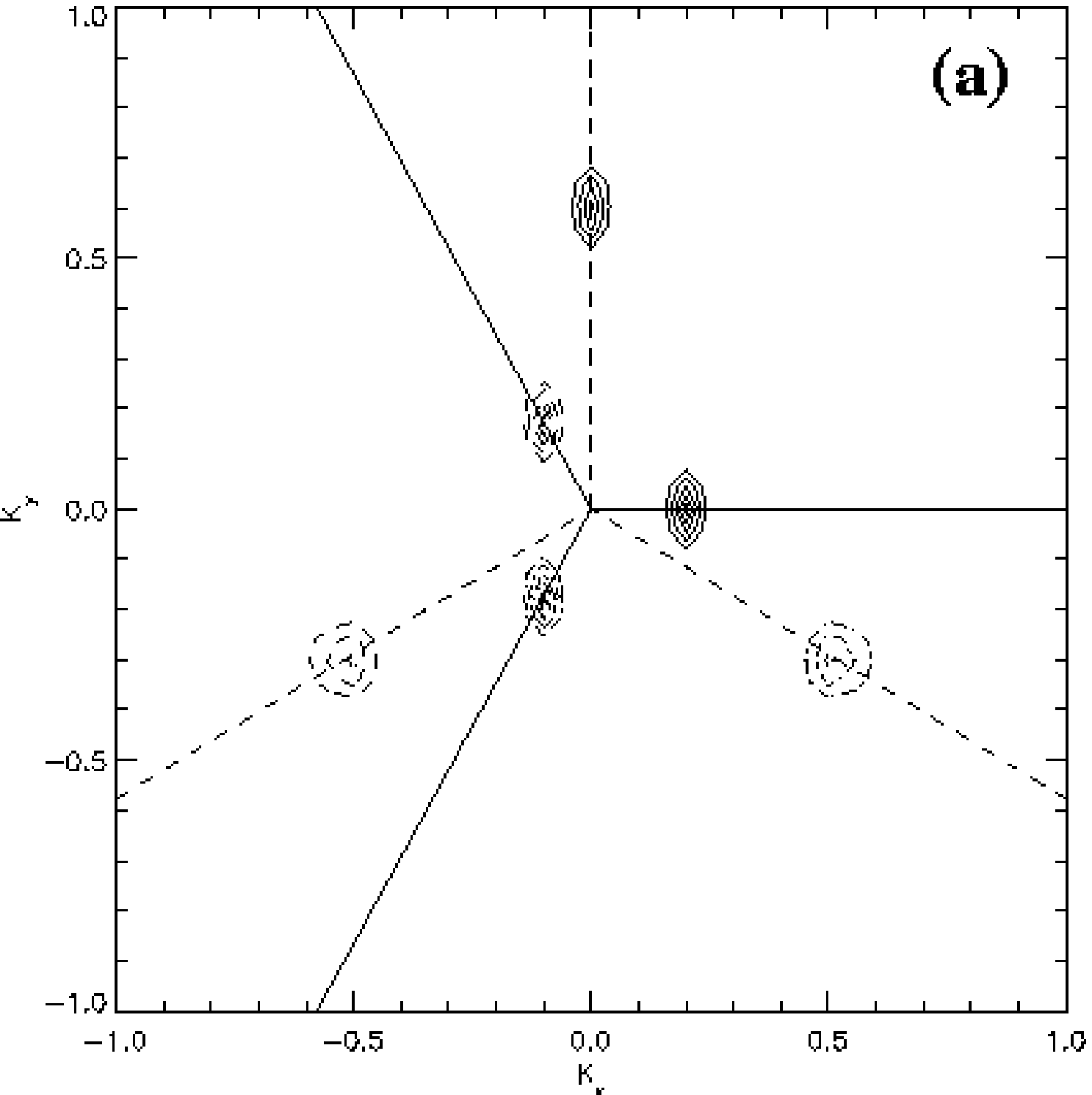}
\hspace{0.2cm}\epsfxsize=5.0cm\epsfbox{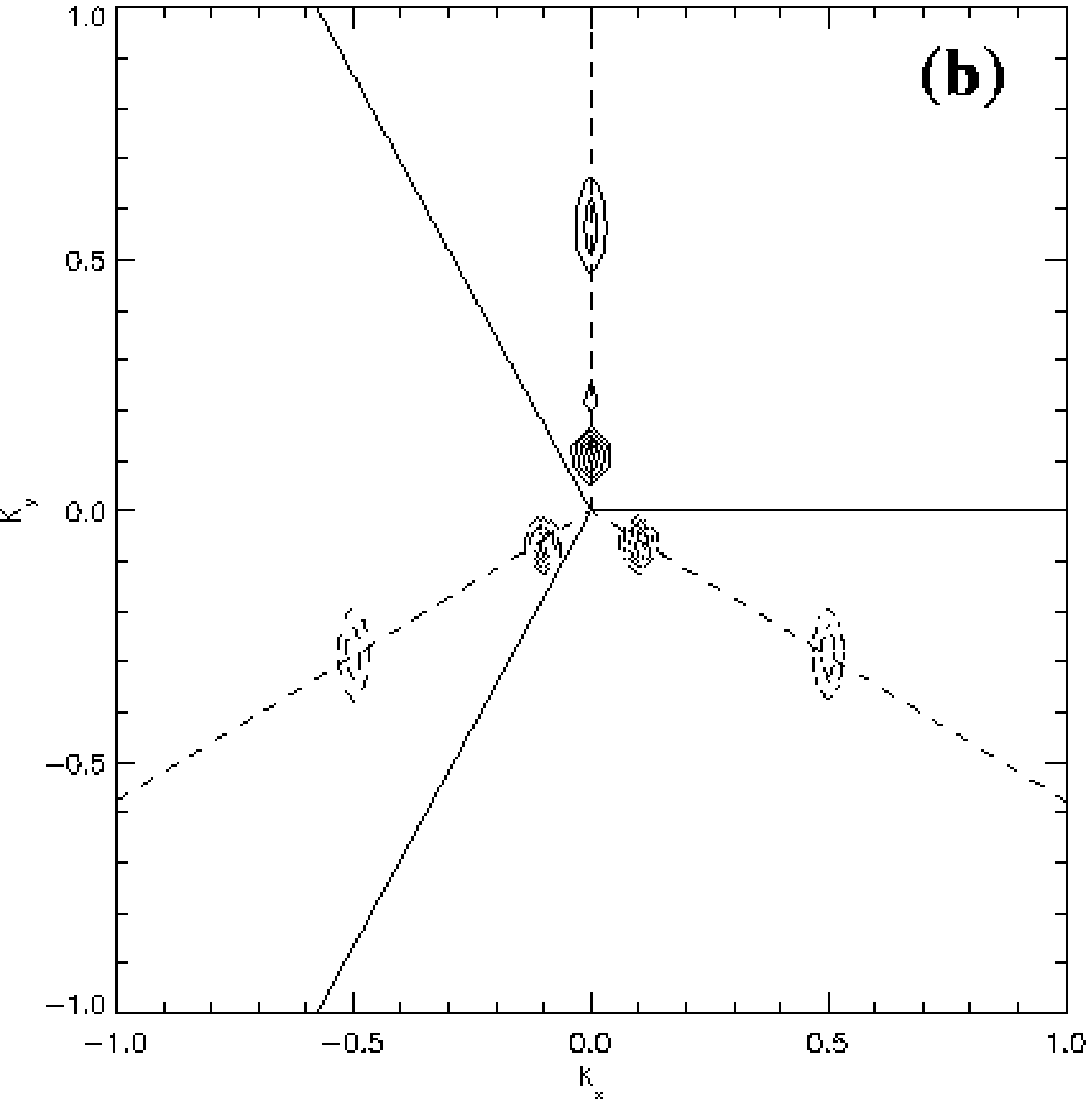}
\hspace{0.2cm}\epsfxsize=5.0cm\epsfbox{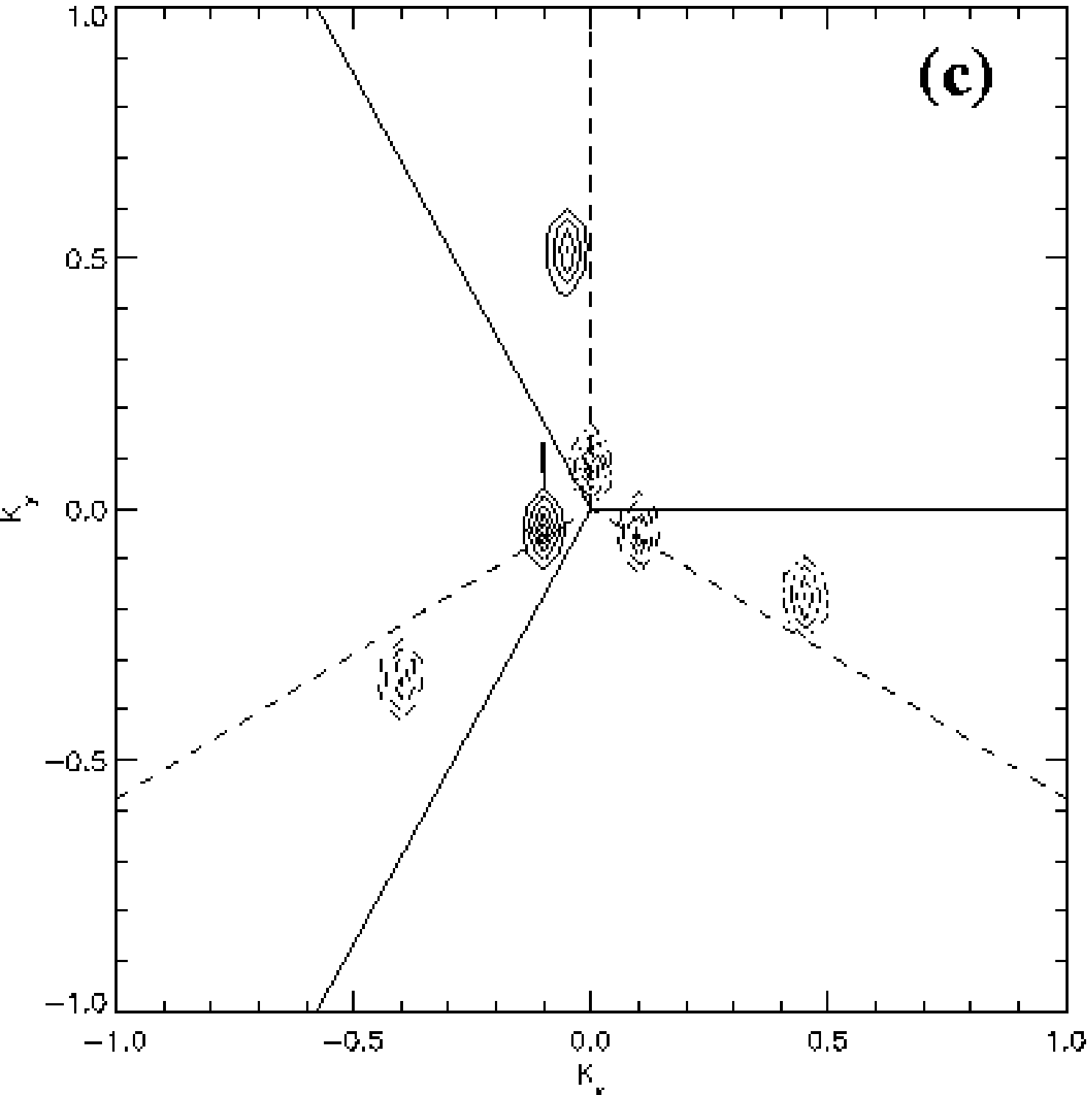}}
\caption{
Wave vector spectra of the complex amplitudes shown in figure
\ref{fig:defects_01}. Panel (a) is for the initial conditions,
and panels (b) and (c) are snapshots at $t=80000$
for $\beta=0$ and $\beta=-2.0$, respectively.
}
\label{fig:defects_01sp}
\end{figure}

To identify the origin of these slow dynamics we performed simulations  starting
from patterns with two straight grain boundaries separating two  hexagon
patterns of different orientation rotated by $\pi/2$ relative to each other.  For
certain magnitudes of the initial wavevectors the grain
boundaries did not annihilate  each other but persisted for an exceedingly
long time. Fig.\ref{fig:defects_01}a shows such an initial condition with
$\vec{q}_2=0.2$. A  contour plot of the histogram of the local wavevector of
each of the  three modes in the initial condition fig.\ref{fig:defects_01}a is
shown in \ref{fig:defects_01sp}a.  Solid lines pertain to $q_1$, dashed
lines to $q_2$, and  dashed-dotted lines to $q_3$. Despite the large
magnitude of the reduced wavevector in the center domain the pattern is still
linearly stable since the longitudinal component of the wavevector  vanishes,
$\vec{q}_j\cdot \hat{n}_j=0$. Within  (\ref{eq:mf1},\ref{eq:mf2}) only this
projection enters the stability  conditions.  
Either without ($\beta=0$) and with ($\beta=-2$) mean flow
the initial  condition evolves to the patterns shown in
figs.\ref{fig:defects_01}b and c,  respectively, for $t=80,000$ and  $\mu=0.5$. 
For $\beta=0$ simulation beyond $t=80,000$ showed little difference in the
spatial structure,  and the defects seem to have reached asymptotic,
immobile states at the end of the simulations.
In the $\beta=-2.0$ case, although the spectra and the domain sizes  
remain more of less the same after $t=80,000$, 
defects exhibit persistent lateral motion along the grain boundaries
(velocity $\sim 4\times 10^{-4}$)
throughout the simulation which continues beyond $t=160,000$ and lead
to a slightly fluctuating shape of the domains.

           
The histograms of the local wavevector of the final states depicted in 
fig.\ref{fig:defects_01sp}(b,c) show that  without mean flow all three
wavevectors are essentially perpendicular to $\hat{n}_i$ implying that
individual defects would not move. In the simulation with mean flow the 
wavevectors are clearly not perpendicular to  $\hat{n}_i$. Instead,
$\vec{q}_2\cdot\hat{n}_2 \approx -0.1$ and 
$\vec{q}_3\cdot\hat{n}_3\approx -0.1$.  In separate simulations we find  that
for $\beta=-2$ and $\mu=0.5$ this is the wavenumber at which  individual
PHD's do not move, $q_{st}=-0.1$. The histogram for the third wavevector
$q_1$, however,  has peaks at $(q_x, q_y) = (-0.1,-0.05)$ and $(-0.05,
0.6)$.  Thus,  the two peaks have different projections onto $\hat{n}_1$ and
only one of them agrees with  $q_{st}$. This may be  related to the fact that
$A_1$ has no dislocations in the grain boundary. In separate simulations of
individual PHD's with dislocations in $A_2$ and $A_3$  we find that the
velocity of the PHD  depends only weakly on the wavenumber of the
defect-free component ($A_1$), and over the whole range  
$-0.15<\vec{q}_1\cdot\hat{n}_1<-0.05$ the PHDs are essentially motionless.
Specifically, the magnitude of the velocity is below $5\times10^{-4}$  for $\mu=0.5$,
$\beta=-2$, $\vec{q}_2\cdot\hat{n}_2=-0.1$, and
$\vec{q}_3\cdot\hat{n}_3=-0.1 $. 

With and without mean flow, the orientation of the initial and  of the  final
wavevectors in the top and  bottom domains are different implying that in
these domains the pattern  rotated until the projections of  their wavevectors
reached $q_{st}$. This suggests that, more generally, in the ordering
dynamics of hexagons starting from random initial conditions the orientation of
a hexagon pattern within a given domain may rotate in a similar fashion and
in the long-time dynamics the orientation of adjacent domains may
predominantly be close to the stationarity condition for PHD's, i.e. the
projections of their wavevectors onto a suitably  chosen $\hat{n}$ have the
same value and that value is close to  $q_{st}$. Of course, since our results 
are based on Ginzburg-Landau equations  they only apply to grain
boundaries across which the orientation  changes only by small amounts.
Furthermore, in the truncation (\ref{eq:mf1},\ref{eq:mf2})  the higher-order
transverse derivatives $\nabla \cdot \hat{\tau}_i$ have  been neglected.
They are expected to modify the stationarity condition,  since the defect
velocity will also depend on the transverse component  of the  wavevector,
$\vec{q}_i\cdot\hat{\tau}_i$. The independence of the  velocity on the
transverse component within (\ref{eq:mf1},\ref{eq:mf2}) is related to  the
isotropy of the system. Thus, it is expected that at higher orders  the
condition $\vec{q}_i\cdot\hat{n}_i=q_{st}$ is replaced by a more 
complicated, but qualitatively similar condition suggesting similar  behavior
for grain boundaries.

\section{Conclusion}
\label{sec:con}

The importance of mean flows has been noted in a wide range of
pattern-forming systems.  Various different types can be distinguished. In
systems like binary-mixture convection \cite{ClKn92} and  in surface waves
on liquids with small viscosity \cite{VeKn01} they are driven by traveling
waves. In other systems they correspond to a  Goldstone mode as  is the
case in free-slip convection (e.g.
\cite{Be94,MaCo00a}) or they arise  from a conservation law, e.g. in systems
with a deformable interface 
\cite{GoNe95a,ReRe93}. Then they constitute a separate dynamical variable and 
satisfy an evolution equation of their own. Depending on the symmetries  of
the system, the additional degree of freedom can introduce a host of new
phenomena (e.g. \cite{MaCo00,RoRi01,RiGr98}). In this paper we have 
studied the mean flow that is driven in Rayleigh-B\'enard convection by
deformations of the convection pattern, which becomes relevant even close
to onset for fluids with low Prandtl number \cite{SiZi81,DePe94,Be94}. In the
realistic  case of no-slip boundary conditions it does not constitute an 
additional dynamical variable.  Due to the incompressibility of the fluid it can
only arise in three-dimensional systems, i.e. in two-dimensionally extended
patterns,  and introduces then a non-local interaction. Experimentally,  its
most striking signatures are the skew-varicose instability and the 
appearance of spiral defect chaos.   The experimental observation of the
skew-varicose  instability and of transients of spiral-defect chaos  in
standing waves in vertically vibrated granular media \cite{BrBi98,BrLe01}
suggests that a  similar mean flow may also be relevant in that system. 
 
The focus of our weakly nonlinear analysis has been the impact of the mean
flow on the  stability and dynamics of hexagonal patterns. We have extended
the usual three coupled Ginzburg-Landau equations for the description of
hexagonal patterns by an  equation for a vertical-vorticity mode in direct
extension of previous work on  roll convection at small Prandtl numbers
\cite{DePe94,Be94}. 

In general, there are two side-band instabilities that limit the band  of stable
wavenumbers of hexagons. In the absence of mean flow it is always the
longitudinal long-wave mode that is the relevant destabilizing mode
immediately above the  saddle-node bifurcation at which the hexagons come
first into  existence,  while the transverse long-wave mode is the relevant
mode for larger amplitudes. The  Rayleigh number for the  
cross-over from one to the other mode depends on the cross-coupling 
coefficient $\nu$, but for realistic value it always occurs very close  to the
saddle-node bifurcation. The longitudinal mode is therefore  poorly
accessible in the absence of mean flow (e.g. \cite{SuTs94}). We found
that in the long-wave limit only one of the two phase modes, the transverse
mode, couples to  the mean flow and consequently only its stability limit
depends on the  Prandtl number. Specifically, for sufficiently small Prandtl 
numbers ($\sim 2.6$ for $\nu=2$) the transverse mode is stabilized for
wavenumbers below the  critical wavenumber to the extent that it is
preempted by the  longitudinal mode over the whole range of Rayleigh
numbers from the saddle-node  bifurcation to the transition to the mixed
mode. In this  regime studies of the difference between the two modes should
be  accessible in experiments  since the type of instability encountered 
depends on whether the stability limits are crossed at the low- or the 
high-$q$ side of the stability balloon.

Our simulations of the nonlinear evolution of the instabilities  indicate  that,
compared to the longitudinal instability, the transverse instability leads to a
considerably larger  number of penta-hepta defects and to more grain
boundaries separating  patches of hexagons that are rotated with respect to
each other. While indications of this were  also found in  the absence of the
mean flow \cite{SuTs94}, the mean flow  makes the distinction clear  enough
to make it worth addressing experimentally. To do so,  hexagon patterns with
a wavenumber away from the band center need to be initiated.  Recently it
has been shown that such initial conditions can, in fact, be prepared  by  a
suitable localized heating of the fluid \cite{Scpriv,BoPepriv}.  Given the
striking difference in the transients arising from the  instabilities of the 
two modes it would be interesting to bring these techniques to bear in 
this system.

Although the coupling to the mean flow makes the system non-variational 
no regime has been identified in which oscillatory instabilities are  relevant.
We also find that the stability limits are always determined  by long-wave
instabilities and no additional instabilities at finite  wavenumbers arise.
Furthermore, to the order considered, the coupling  to the mean flow
vanishes at the band center. This implies that the  pattern is always linearly
stable there. 

 As is the case in roll convection, the mean flow also affects the  motion of
defects. For the penta-hepta defects relevant in hexagonal  patterns we find
that similar to the case of rolls the wavenumber at  which the defect is
stationary is shifted to wavenumbers smaller than  the critical one. For
coarsening experiments starting from random  initial conditions one may
therefore expect that the eventual  wavenumber of the ordered pattern may
be reduced correspondingly. Our simulation suggest that the dependence  of
the defect velocity on the wave vector allows one to predict which grain
boundaries have a particularly long life time.  
Furthermore, a persistent drift of the defects
in the grain boundary is also observed in the simulations. 
One may also expect that similar to the case of dislocations 
in roll  patterns \cite{Bopriv} the mean flow
may allow two penta-hepta  defects to form stable pairs if the background
wavenumber is between  the critical wavenumber and that corresponding to
stationary defects.

We gratefully acknowledge useful discussions with  G. Ahlers,  E.
Bodenschatz, B. Echebarria, D. Egolf and S. Venkataramani. This work is
supported by NASA (NAG3-2113) and the Engineering Research Program of
the Office of Basic Energy Sciences at the Department of Energy
(DE-FG02-92ER14303). YY acknowledges computation support from the
Argonne National Labs and the DOE-funded ASCI/FLASH Center at the
University of Chicago.

\appendix
\section{Derivation of the nonlinear phase equations}
\label{appendA}
Here we derive the nonlinear phase equations at the codimension-two point
$(q^{(ct)}, R^{(ct)})$  where both $\sigma_t$ and $\sigma_l$ are zero. Mainly,
to obtain  explicit expressions for $(q^{(ct)}, R^{(ct)})$ we consider weak
mean flow, $\beta \ll 1$. (cf. eq.(\ref{eq:codim2_beta})).   We rescale $X=\delta
x$, $Y=\delta y$, and $T=\delta^4 t$. In this expansion $\delta$ and $\beta$
are two independent small parameters. Here we expand the amplitudes 
$A_j=r_j e^{i q \hat{n}_j\cdot (x,y)+i \phi_j}$ as
\begin{eqnarray}
\label{eq:appen1}
r_j &=&R^{(ct)}_0+\delta^2 r_{j2} + \delta^4 r_{j4}+\cdots,\\
\phi_j&=&\delta(\phi_{j0}+\delta^2\phi_{j2}+\delta^4\phi_{j4}+\cdots),\;\;\; j=1,\;2,\;3,
\end{eqnarray}
with $R^{(ct)}_0$ given in eq.(\ref{eq:codim2_beta}).
The mean flow amplitude $Q$ is expanded accordingly in $\delta^2$
\begin{eqnarray}
\label{eq:appen2}
Q&=&\delta^2 Q_2+\delta^4 Q_4+\cdots.
\end{eqnarray} 

We substitute the above expansions into eqs.(\ref{eq:mf1},\ref{eq:mf2})  and
solve them at successive orders for $\delta$. The mean flow feeds back to
the equations for the phases $\phi_j$ {\em via} the amplitude modulations (cf.
eq.(\ref{eq:mf2})). Thus, at each order we first solve for the  amplitudes and
the mean flow in terms of the phases that were determined at the  previous
orders, and substitute these solutions  into the phase equations to  obtain the
phases at the next order.  
At ${\cal O}(\delta^2)$, up to first order in $\beta$, 
\begin{eqnarray}
\label{eq:appen3}
r_{12}  &=& -\frac{1}{\sqrt{2(1+\nu)}}\partial_x\phi_x
         -\frac{3}{8\nu(1+\nu)}(\partial_x\phi_x-3\partial_y\phi_y)\beta,\\
r_{22}  &=& -\frac{1}{4\sqrt{2(1+\nu)}}
            (\partial_x-\sqrt{3}\partial_y)(\phi_x-\sqrt{3}\phi_y)\nonumber\\
      && +\frac{3}{8\nu(1+\nu)}\bigl[(2\partial_x+\sqrt{3}\partial_y)\phi_x+\sqrt{3}\partial_x\phi_y\bigr]\beta,\\
r_{23}  &=& -\frac{1}{4\sqrt{2(1+\nu)}}
            (\partial_x+\sqrt{3}\partial_y)(\phi_x+\sqrt{3}\phi_y)\nonumber\\
      && +\frac{3}{8\nu(1+\nu)}\bigl[(2\partial_x-\sqrt{3}\partial_y)\phi_x-\sqrt{3}\partial_x\phi_y\bigr]\beta,\\
Q_2    &=& -\left(\frac{3}{4\nu\sqrt{2(1+\nu)}}+
           \frac{9(3\nu+1)}{16\nu^3(1+\nu)}\beta\right)(-\hat{e}_z\cdot\nabla\times\vec{\phi}).
\end{eqnarray}
 At cubic order we recover eq.(\ref{eq:phase}). 
In order to go on to fifth order we require that both
$\sigma_l$ and $\sigma_t$  vanish (up to second order in $\beta$). As in
section \ref{sec:linear01}, the solutions are expressed in terms of the 
translation modes $\phi_x=-(\phi_{20}+\phi_{30})$ and
$\phi_y=(\phi_{20}-\phi_{30})/\sqrt{3}$.

Repeating the same procedures at ${\cal O}(\delta^4)$, we obtain $r_{j4}$.
The expressions are too long to be displayed here.  Also, at this order it is 
impossible to solve for $Q_4$ in closed form.  We  therefore take the
Laplacian of the phase equations at this order and substitute $r_{j4}$ and
$\nabla^2 Q_4$ to obtain the nonlinear equations for $\phi_x$ and
$\phi_y$ at the codimension-two point:

\begin{eqnarray}
\label{eq:appen4}
\partial_t \nabla^2\phi_x &=& {\cal L}_x + \mbox{NL}_{x0}+ \beta \mbox{NL}_{x1}, \\
\partial_t \nabla^2\phi_y &=& {\cal L}_y + \mbox{NL}_{y0}+ \beta \mbox{NL}_{y1},
\end{eqnarray}

where ${\cal L}_x$ and ${\cal L}_y$ are the linear terms given on the 
right-hand-sides of eq.(\ref{eq:phase:1}) and eq.(\ref{eq:phase:2}), respectively.
The nonlinear terms at zeroth order in $\beta$ for both equations, denoted as
$\mbox{NL}_{x0}$ and $\mbox{NL}_{y0}$, read as follows:
\begin{eqnarray}
\label{eq:appen5}
\mbox{NL}_{x0} =&&-\frac{q_0\nu^2}{(1+\nu)^2} \nabla^2\Bigg[
\partial_x\phi_x\left(\frac{33+25\nu}{2}\partial_x^2\phi_x+
                      \frac{3+5\nu}{2}\partial_y^2\phi_x+(3+8\nu)\partial_{xy}^2\phi_y\right)+
\nonumber\\
&&(\partial_y\phi_x+\partial_x\phi_y)\left((3+5\nu)\partial_{xy}^2\phi_x+\frac{3+5\nu}{2}\partial_x^2\phi_y+
                      \frac{3(3+\nu)}{2}\partial_y^2\phi_y \right)
+\nonumber\\
&&\partial_y\phi_y\left(\frac{3(3+\nu)}{2}\partial_{y}^2\phi_x+\frac{3+11\nu}{2}\partial_x^2\phi_x+
                      3(3+2\nu)\partial_{xy}^2\phi_y \right) \Bigg],
\end{eqnarray}
\begin{eqnarray}
\label{eq:appen6}
\mbox{NL}_{y0} =&&-\frac{q_0\nu^2}{(1+\nu)^2} \nabla^2\Bigg[
\partial_x\phi_x\left(\frac{3+8\nu}{2}\partial_{xy}^2\phi_x+
                      \frac{3+5\nu}{2}\partial_x^2\phi_Y+\frac{9(1+\nu)}{2}\partial_{y}^2\phi_y\right)+
\nonumber\\
&&(\partial_y\phi_x+\partial_x\phi_y)\left(\frac{3+5\nu}{2}\partial_{x}^2\phi_x+\frac{3(3+\nu)}{2}\partial_y^2\phi_x+
                      3(3+\nu)\partial_{xy}^2\phi_y \right)
+\nonumber\\
&&\partial_y\phi_y\left(3(3+2\nu))\partial_{xy}^2\phi_x+\frac{3(3+\nu)}{2}\partial_x^2\phi_y+
                      \frac{27(1+\nu)}{2}\partial_{y}^2\phi_y \right) \Bigg],
\end{eqnarray}                       
with $q_0$ defined in eq.(\ref{eq:codim2_beta}).
The expressions for $\mbox{NL}_{x1}$ and $\mbox{NL}_{y1}$ are 
too long to be shown here.  

%

\end{document}